\documentclass[usenatbib]{mnras}

\usepackage[T1]{fontenc}
\usepackage{ae,aecompl}

\usepackage{graphicx}	
\usepackage{amsmath}	
\usepackage{amssymb}	
\usepackage{booktabs}
\usepackage{eso-pic}

\usepackage{ulem}       
\usepackage{url}
\usepackage{times}
\usepackage{array}
\usepackage{dirtytalk}
\usepackage{verbatim}
\usepackage{gensymb}

\def \cf2{{\it Cosmicflows-2\,}}

\def\hi{\textsc{HI~}}

\def\mnras{Monthly Notices of the Royal Astronomical Society}
\def\apj{The Astrophysical Journal}
\def\apjl{Astrophysical Journal Letters}
\def\aap{Astronomy and Astrophysics}

\def\prd{Physical Review D}

\def\apjs{The Astrophysical Journal Supplement Series}
\def\aj{The Astronomical Journal}
\def\pasa{Publications of the Astron. Soc. of Australia}

\makeatletter
\let\ftype@table\ftype@figure
\makeatother

\title[PCA method with FAST]{Recovering 21-cm signal from simulated FAST intensity maps}
\author[Yohana et al.]{Elimboto Yohana$^{1,2,3,4}$, Yin-Zhe Ma$^{3,4 \dagger}$,
Di Li$^{5,6,4}$, 
Xuelei Chen$^{7,6,8}$, 
Wei-Ming Dai$^{3,4}$
\\
$^{1}$ Astrophysics and Cosmology Research Unit, School of Mathematics, Statistics \& Computer Science, University of KwaZulu-Natal, \\ Westville Campus, Private Bag X54001, Durban, 4000, South Africa \\
$^{2}$ Dar Es Salaam University College of Education, A Constituent College of the University of Dar Es Salaam, P.O. Box 2329 Dar Es Salaam, Tanzania \\
$^{3}$ Astrophysics and Cosmology Research Unit, School of Chemistry and Physics, University of KwaZulu-Natal, Westville Campus, \\ Private Bag X54001, Durban, 4000, South Africa \\
$^{4}$ NAOC-UKZN Computational Astrophysics Centre (NUCAC), University of KwaZulu-Natal, Durban, 4000, South Africa\\
$^{5}$ CAS Key Laboratory of FAST, National Astronomical Observatories, Chinese Academy of Sciences, Beijing 100101, China\\
$^{6}$ School of Astronomy and Space Sciences, University of Chinese Academy of Sciences, Beijing 100049, China\\
$^{7}$ Key Laboratory of Computational Astrophysics, National Astronomical Observatories, Chinese Academy of Sciences, Beijing 100012, China \\
$^{8}$ Centre for High Energy Physics, Peking University, Beijing 100871, China \\
{\rm $^{\dagger}$Corresponding author}: ma@ukzn.ac.za  \\
}

\pubyear{2018}

\begin{document}
\label{firstpage}
\pagerange{\pageref{firstpage}--\pageref{lastpage}}
\maketitle

\begin{abstract}
The 21-cm intensity mapping (IM) of neutral hydrogen (\hi\!\!) is a promising tool to probe the large-scale structures. Sky maps of 21-cm intensities can be highly contaminated by different foregrounds, such as Galactic synchrotron radiation, free-free emission, extragalactic point sources, and atmospheric noise. We here present a model of foreground components and a method of removal, especially to quantify the potential of Five-hundred-meter Aperture Spherical radio Telescope (FAST) for measuring HI IM. We consider 1-year observational time with the survey area of $20,000\,{\rm deg}^{2}$ to capture significant variations of the foregrounds across both the sky position and angular scales relative to the \hi signal. We first simulate the observational sky and then employ the Principal Component Analysis (PCA) foreground separation technique. We show that by including different foregrounds, thermal and $1/f$ noises, the value of the standard deviation between reconstructed 21-cm IM map and the input pure 21-cm signal is $\Delta T = 0.034\,{\rm mK}$, which is well under control. The eigenmode-based analysis shows that the underlying \hi eigenmode is just less than $1$ per cent level of the total sky components. By subtracting the PCA cleaned foreground+noise map from the total map, we show that PCA method can recover \hi power spectra for FAST with high accuracy.

\end{abstract}
%

%



\section{Introduction}
\label{sec:intro}

Large-scale structures of the Universe can be efficiently surveyed by the neutral hydrogen (\hi\!\!) intensity mapping (IM) technique, which measures the 21-cm emission line of neutral atomic hydrogen (\hi\!\!). The \hi IM technique is a promising approach to measure the collective \hi emission intensity over the physical volume of a few tens of Mpc, to efficiently survey massive amounts of galaxies without resolving individual sources~\citep{21cm-review, Battye_2013, Bull_2015, Kovetz_2017}. Although the 21-cm emission signal is weak, observations over a large number of sky pixels through IM can significantly enhance the collective \hi detection sensitivity. With \hi IM, we take advantage of the single large dish (in particular FAST) which generally has better absolute gain and can sample the fluctuations over large angular scales.

Several near-term and future radio experiments aim to use \hi IM technique to probe the large-scale structure of the Universe and constrain cosmological parameters.  In our series of intensity-mapping with \hi studies, we have prioritized to work with some of such single-dish radio telescopes, namely; FAST~\citep{Nan_2011, Li_2016, Li_2018},
BINGO~\citep{Battye_2012, Battye_2013, 2014arXiv1405.7936D,Bigot-Sazy_2015, Battye_2016}, MeerKLASS~\citep{MeerKLASS}; and SKA-MID~\citep{SKAI2, SKAI, SKA_science, SKAI_Red}
in a single-dish imaging mode~\citep{Yohana_2019}. For instance, FAST can offer a high resolving power since it is currently the largest single-dish telescope in the world~\citep{Peng_2009}. Being a medium-sized telescope with special design~\citep{Battye_2016}, BINGO is optimized to detect the Baryon Acoustic Oscillations (BAO) at radio frequencies, which would, in turn, be useful to measure the dark energy equation of the state. SKA-MID telescope array is suitably optimized to probe cosmological scales, large volume of the Universe. These next-generation experiments for large-scale structure surveys of the Universe are suitable laboratories to learn various \hi IM techniques. In this study, we will focus on FAST, which has already been commissioned for initial tests, and the prior data for $20$ hours of integration time is already available. Therefore, we intend to focus the FAST capability of delivering 21-cm intensity data by simulating mock sky signal and foregrounds.

IM approach is promising, but the method shifts the observational problem from that of weak \hi detection to that of foreground contamination. The performance of \hi IM surveys in detecting and extracting \hi signal will, therefore, depend on the successful removal of foregrounds and other contaminants, calibration of instruments and mitigation of several problems on the large scales~\citep{Pourtsidou_2017}. Luckily, total foreground contaminants should have a smooth frequency dependence~\citep{Liu_and_Tegmark_2011,  Alonso_D_2015, Bigot-Sazy_2015, GNILC, Villaescusa-Navarro, Cunnington_2018}, whereas the underlying 21-cm signal varies in frequency and sky position. Property of smoothness means that foreground modes are correlated in frequency~\citep{Santos_2005} hence can be clustered in the direction of maximum variance and stripped out by appropriate methods. But noise and systematics are expected to be spectrally uncorrelated, except for the correlated $1/f$ noise~\citep{Harper_2018}.

Many approaches to address the foreground cleaning have been tested and presented in the works of literature so far. These include the line-of-sight fitting method~\citep{Line-of-sight, Liu_and_Tegmark_2011}, line-of-sight and Wiener filter~\citep{line_of_sight2}, and the method of foregrounds signal frequency cross-correlation~\citep{Santos_2005}. More recently, Robust Principal Component Analysis (RPCA)~\citep{RPCA}, Independent Component Analysis (ICA) techniques~\citep{blind_ICA, fast_ICA, FASTICA, FASTICA2, Alonso_D_2015}, extended ICA~\citep{2016ApJS..222....3Z}, Singular Value Decomposition (SVD)~\citep{SVD1, Masui_2013}, correlated component analysis (CCA)~\citep{Bonaldi_2006}, Principal Component Analysis (PCA)~\citep{GBT, Villaescusa-Navarro, Bigot-Sazy_2015, Alonso_D_2015} and methods that assume some physical properties of the foregrounds, such as polynomial/parametric-fitting~\citep{Bigot-Sazy_2015, Alonso_D_2015} have been widely deployed. Other approaches, for example, quadratic estimation~\citep{Switzer_2015} and inverse variance
\citep{Liu_and_Tegmark_2011} are also being discussed and investigated. These foreground contaminant subtraction algorithms are successful to some extent, but still have issues, such as biased results and the inability to mitigate various systematics.
For example, FASTICA~\citep{blind_ICA, fast_ICA} seems to succeed in removing dominant foreground contaminants, especially, resolved point sources and diffuse frequency-dependent components on large scales, but fails to mitigate systematics on smaller scales dominated by thermal noise~\citep{FASTICA2}.

This work investigates the potential of \hi IM FAST studies and the validity of the foreground removal through, particularly, the PCA analysis. We will simulate the 21-cm sky and various foregrounds using FAST telescope parameter specifications, and apply the PCA foreground cleaning technique to the map. Although the PCA approach is a general dimensionality reduction and a component separation approach to subtract foregrounds for various contaminated models, each experiment is unique in its specification so how it works for FAST is worth investigating. At the time of writing this manuscript, a similar but different study of forecasting \hi galaxy power spectrum and IM are conducted in~\citet{Wenkai_Hu_2019}. \citet{Wenkai_Hu_2019} made a simulation-based foreground impact study on the measurements of the 21-cm power spectrum with FAST and calculated the expected cosmological parameter precision based on the Fisher matrix with Gaussian instrumental noise. In this paper, we plan to take the foreground problem with FAST IM observations further by adding a complete package of foreground contaminants and correlated $1/f$ noise and challenging the foreground removal method. With more detailed and sophisticated input of \hi IM foreground and instrumental noise, as well as considering a wide FAST sky strip, our approach is a ``closer to reality'' forecast for FAST \hi IM study.

This paper is organized as follows. In Section~\ref{FAST_telescope}, we briefly review the FAST telescope, focusing on its experimental and observational prospects. In Section~\ref{foreground_and_noises}, we discuss the signal and foreground contaminants of the radio sky and the instrumental noises. Section~\ref{PCA_algorithm} is dedicated to the qualitative and quantitative description of the principal component analysis algorithm used for component separation. Section~\ref{HI_recovery} presents PCA results and analyses the recovered 21-cm signal versus the input. We conclude in the last section.

Throughout the paper, while computing the theoretical 21-cm power spectra at different frequencies, we adopt a spatially-flat $\Lambda$CDM cosmology model with best-fitting parameters fixed to {\it Planck} 2013 results, i.e. $\Omega_{\rm b}h^{2}=0.02205$, $\Omega_{\rm c}h^{2}=0.1199$, $n_{\rm s}=0.9603$, and $\ln(10^{10}A_{\rm s})=3.089$~\citep{Planck-collaboration}.

\section{FAST telescope}
\label{FAST_telescope}
Five-hundred-metre Aperture Spherical radio Telescope (FAST)~\citep{Peng_2009, Nan_2011, Li_2016, Li2019} is a multi-beam radio telescope potentially suitable for 21-cm IM surveys. The construction was completed in 2016, and the commissioning phase is drawing to an end. This telescope can map the large-scale cosmic structures and deliver the redshifted 21-cm sky intensity of temperature maps over a wide range of redshifts. Using simple drift-scan (preferred for better spatial sampling) designated as a Commensal Radio Astronomy FasT Survey (CRAFTS)~\citep{Li_2018}, and a transverse set of beams, FAST can survey a broad strip of the sky. With CRAFTS observations using an L-band array of $19$ feed-horns (and $1.05 - 1.45$ GHz), data from different pointings or beams can be combined to construct a high-quality \hi image. In terms of sensitivity, FAST will be more sensitive within its frequency band than any single-dish telescope; its design and features supersedes the $300$-meter post-Gregorian upgrade Arecibo Telescope and
$100$-meter Green Bank Telescope (GBT). FAST has approximately twice and ten times, respectively, the effective collecting areas of Arecibo and GBT~\citep{Li_2016}, and will deliver $10\%$ of the SKA collecting area~\citep{Li_2018}.

In Table~\ref{inst_and_survey_params}, we list all the essential instrumental parameters of the current FAST telescope. Here we consider a survey by FAST conducted in the drift scan mode, which is operationally simple and stable, and works more efficiently for large sky coverage. We consider a survey similar to those presented in the CRAFTS proposal, which will scan a $26'$ wide strip along the Right Ascension direction for each sidereal
day, expected to cover the northern/FAST sky between $-14^{\circ}$ and  $+65^{\circ}$ of declination in about $220$ full days~\citep{Li_2018}. We refer the interested readers for more details of FAST technical designs, survey strategies, capacities, and science potentials in~\citet{Nan_2011},~\citet{Li_2016} and~\citet{Li_2018}.


\begin{table*}
\begin{center}
\caption{FAST instrumental and survey Parameters. The L-band sensitivity is defined as the effective antenna area per system noise temperature. The zenith angle (sky coverage) has the full gain at $26.4^{\circ}$ and (maximum) $18\%$ gain loss at the $40^{\circ}$.}
\label{inst_and_survey_params}
\begin{tabular}{>{\raggedright\arraybackslash}m{50mm}|>{\raggedright\arraybackslash}m{33mm}|>{\raggedright\arraybackslash}m{70mm}}\hline\hline
Parameter description & Value & Reference\\\hline
\multicolumn{3}{c}{Instrumental Parameters} \\\hline
Dish/aperture diameter & $500$\, m & \citet{Nan_2011, Bigot-Sazy_2016, Li_2018} \\
illuminated aperture & $D=300$ m & \citet{Nan_2011, Bigot-Sazy_2016, Li_2018} \\
Frequency coverage & $\nu=1,050$ -- $1,450$\,MHz & \citet{Nan_2011, Bigot-Sazy_2016, Li_2018}\\
Survey redshift range& $0<z<0.35$ & $z=(1420\,{\rm MHz}/\nu) -1$ \\
System temperature & $T_{\rm sys} = $ $20$ K & \citet{Nan_2011,Li_2016,Wenkai_Hu_2019} \\
Number of L-band receivers & $n_{\rm f} = 19$ & \citet{Nan_2011, Bigot-Sazy_2016, Li_2018} \\
L-band sensitivity ($A_{\rm eff}/T_{\rm sys}$)  & $ 1,600 - 2,000\,{\rm m}^{2}\,{\rm K}^{-1}$ 
& \citet{Nan_2011, Li_2016, Li_2018}\\
Telescope positions [latitude, longitude] & North $25^{\circ}48'$, East $107^{\circ} 21'$ 
& \citet{Li_2016} \\
FWHM at reference frequency ($1420$ MHz) & $2.94$ arcmin & \citet{Nan_2011,Li_2016, Li_2018}  \\
Frequency bandwidth (Number of channels)  & $\Delta \nu = 10$ MHz $(N_{\nu} = 40)$ & This paper's choice \\
\hline
\multicolumn{3}{c}{Survey Parameters} \\\hline
Sky coverage   & $\Omega_{\rm sur} = 20,000 \ {\rm deg}^{2}$ & \citet{Wenkai_Hu_2019} \\
Total integration time& $1$ year & Assumed in this paper \\
Opening angle & $100^{\circ} -  120^{\circ}$ ($112.8^{\circ}$) & \citet{Nan_2011,Smoot_2017}\\
Zenith angle (sky coverage) & $26.4^{\circ} - 40^{\circ}$ & \citet{Nan_2011,Li_2016,Wenkai_Hu_2019} \\
Declination & $-14^{\circ}12' - 65^{\circ} 48'$ & \citet{Li_2016}\\
Pointing accuracy & $8$ arcsec & \citet{Nan_2011,Li_2016,Li_2018}\\
Tracking range & $4-6$ hours & \citet{Nan_2011,Smoot_2017}\\
\hline\hline
\end{tabular}
\end{center}
\end{table*}

\section{Signal, noise and foreground}
\label{foreground_and_noises}

\begin{figure}
\includegraphics[width=8.5cm]{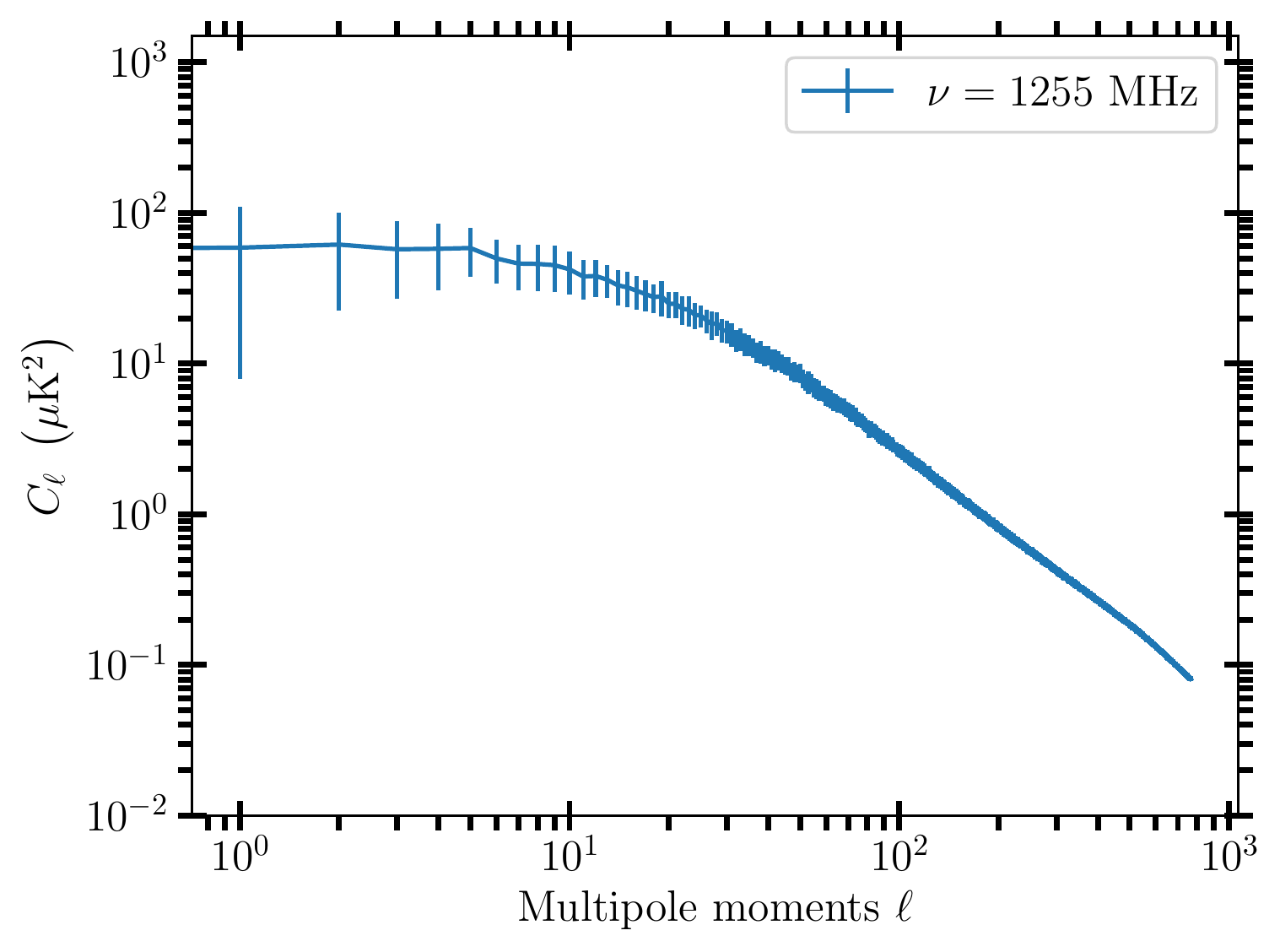}
\caption{The averaged \hi signal power spectrum at the frequency $1255\,{\rm MHz}$ (median frequency of $1050 - 1450$ MHz), and its intrinsic dispersion $\Delta C_{\ell}=\sqrt{M_{\ell\ell}}$ calculated via Eqs.~(\ref{cl_error_cov_matrix}) and (\ref{eq:cl-avg}).}
\label{fig:power_spectra}
\end{figure}

\subsection{\hi Signal}

\subsubsection{\hi brightness temperature}

The observed effective \hi signal brightness temperature is~\citep{Bull_2015, Smoot_2017}
\begin{align}
 T_{\rm b} = \overline{T}_{\rm b}(1 + \delta_{\rm HI}),
\end{align}
which consists of homogeneous and fluctuating parts, for which the fluctuating part in a voxel (an individual volume element) is given by
\begin{align}
 \delta T^{\rm S}({\boldsymbol{\theta}}_{i}, \nu_{i}) = \overline{T}_{\rm b}(z)\delta_{\rm HI}(\mathbf{r}_{i}, z),
\end{align}
where
\begin{align}\label{mean_T_b}
 \overline{T}_{\rm b}(z) = \frac{3}{32\pi}\frac{h_{\rm p}c^{3}A_{10}}{k_{\rm B}m_{\rm p}^{2}\nu_{10}^{2}}\frac{(1+z)^{2}}{H(z)}\Omega_{\rm HI}(z)\rho_{{\rm c},0}
\end{align}
is the mean brightness temperature. Here, $i$ labels the volume element (voxel) given by a 2-dimensional angular direction, $\boldsymbol{\theta}_{i}$, and frequency $\nu_{i}$~\citep{Bull_2015}; $\mathbf{r}_{i}$ is the comoving distance to the voxel $i$. $h_{\rm p}$ is the Planck constant, $m_{\rm p}$ is the mass of the proton, $k_{\rm B}$ is the Boltzmann constant, and $c$ is the speed of light. $A_{10} \approx 2.869 \times 10^{-15} \ {\rm s}^{-1}$ is the Einstein's coefficient for spontaneous emission, which is a measure of probability per unit time that a photon with an energy $E_{1} - E_{0} = h_{\rm p}\nu$ is emitted by an electron in state $1$ with energy $E_{1}$, decaying spontaneously to state $0$ with energy $E_{0}$; $\nu_{10}\simeq 1420\,{\rm MHz}$ is the 21-cm rest-frame emission frequency, $H(z)$ is the Hubble rate as a function of redshift, and
\begin{align}
 \Omega_{\rm HI} = \frac{\rho_{\rm HI}}{\rho_{{\rm c},0}}
\end{align}
is the neutral hydrogen fraction such that $\rho_{{\rm c}, 0} = 3H_{0}^{2}/8\pi G$ is the critical density today.
Equation~(\ref{mean_T_b}) can be further simplified to an expression related to cosmological parameters~\citep{Battye_2013, Hall_2013, Bigot-Sazy_2016},
\begin{eqnarray}
 \overline{T}_{\rm b}(z) &=& 0.188{\rm K}\left( \Omega_{\rm HI}(z)h  \right)\frac{(1+z)^{2}}{E(z)} \nonumber \\
              &=& 0.127\left(\frac{h}{0.7}\right)\left(\frac{\Omega_{\rm HI}(z)}{10^{-3}}  \right)\left(\frac{(1+z)^{2}}{E(z)}\right) \, {\rm mK},
\end{eqnarray}
where $h = H_{0}/(100\,{\rm km}\,{\rm s}^{-1}{\rm Mpc}^{-1})$ is the reduced Hubble parameter and $E(z) =H(z)/H_{0}= \sqrt{\Omega_{\rm m} (1+z)^{3} + \Omega_{\Lambda}}$ is the redshift-dependent part of Hubble parameter.

\subsubsection{Power spectrum}

Since most of the \hi are locked within galaxies in the low-redshift universe, it is expected that \hi signal (21-cm brightness temperature) will be a biased tracer of underlying matter fluctuations, naturally characterized by the angular power
spectrum~\citep{Lewis_2007, Datta_2007}. Thus the \hi density contrast, $\delta_{\rm HI}$ is expressed as a convolution of the \hi bias, $b_{\rm HI}$ and the total matter density perturbation $\delta_{\rm m}$:
\begin{align}
 \delta_{\rm HI} = b_{\rm HI} * \delta_{\rm m},
\end{align}
where $\delta_{\rm m}$ is the matter density contrast.

Assuming the peculiar velocity $v$ gradients and $v/c$ terms are small for these pixels (which are in practice large~\citep{Bull_2015, Smoot_2017}), the temperature fluctuation for a given
frequency $\nu$, a solid angle $\Delta \Omega$ of a considered spatial volume element and a frequency interval $\Delta \nu$ is
\begin{align}
 T_{\rm b}(\nu, \Delta \Omega, \Delta \nu) \approx \overline{T}_{\rm b}(z)\left[1 + b_{\rm HI}\delta_{\rm m}(z) - \frac{1}{H(z)}\frac{{\rm d} v}{{\rm d}s} \right],
\end{align}
where ${\rm d}v/{\rm d}s$ is the proper gradient of the perpendicular velocity along the line of sight, which accounts
for the peculiar velocity effect, and $\Omega_{\rm HI}(z) = (1+z)^{-3}\rho_{\rm HI}(z)/\rho_{{\rm c},0}$ is the \hi fractional density. We then use these relations to
calculate in a more consistent manner the redshift evolution of \hi density, 21-cm brightness temperature, and \hi bias, see also~\citet{Bull_2015} and~\citet{Smoot_2017}.

Instead of carrying out exact calculations of the angular power spectrum, 21-cm cosmological signal simulations are performed by generating Gaussian realizations
from the flat-sky approximation of the angular power spectrum (accurate within $1\%$ level for $\ell> 10$, see~\cite{Datta_2007}),
\begin{align}
 C_{\ell}(z, z') = \frac{1}{\pi \chi \chi'}\int_{0}^{\infty}{\rm d}k_{\parallel}{\rm cos}\left( k_{\parallel}\Delta \chi \right)P_{T_{\rm b}}({\bf k}; z, z'),
\end{align}
where $\chi$ and $\chi'$ are comoving distances to redshifts $z$ and $z'$, $\Delta \chi = \chi - \chi'$, and ${\bf k}$ is the vector with components $k_{\parallel}$ and $\ell/\bar{\chi}$, in the direction
parallel and perpendicular to the line of sight respectively~\citep{Shaw_2014, Bigot-Sazy_2015}.
\begin{align}\label{T_b_power_spectrum}
 P_{T_{\rm b}}({\bf k}; z, z') = \overline{T}_{\rm b}(z)\overline{T}_{\rm b}(z')\left( b + f\mu^{2}\right)^{2}P_{\rm m}(k; z, z')
\end{align}
is the 3D power spectrum of the 21-cm brightness temperature, where $\mu \sim k_{\parallel}/k$.

In the definition of terms from Eq.~(\ref{T_b_power_spectrum}), $b$ is the bias which is unity on large scales,
\begin{align}
 f = \frac{{\rm d}\log D}{{\rm d}\log a},
\end{align}
is the linear growth rate, where $D(z)$ is the growth factor. Finally,
the real-space matter power spectrum is given by~\citep{Shaw_2014, Bull_2015, Bigot-Sazy_2015}
\begin{align}
P_{\rm m}(k; z, z') = P(k)D(z)D(z').
\end{align}

To calculate the covariance matrix of the angular power spectrum ($C_{\ell}$)
through simulations, we calculate the 3-dimensional angular power
spectra of 21-cm tomography error covariance matrix
\begin{align}
M = \langle n n^{t} \rangle,
\end{align}
over $N = 100$ samples of \hi sky map realizations.
We compute this covariance matrix by
\begin{eqnarray}
M_{\ell \ell^{'}} = \frac{1}{N}\sum_{i = 1}^{N}\left(C_{\ell}^{(i)} - \overline{C}_{\ell}\right)\left(C_{\ell^{'}}^{(i)} - \overline{C}_{\ell^{'}}\right), \label{cl_error_cov_matrix}
\end{eqnarray}
where $C_{\ell}$ averaging is performed over a particular frequency for all simulated maps:
\begin{eqnarray}
{\overline{C}_{\ell}} = \frac{1}{N}\sum_{i = 1}^{N} C_{\ell}^{(i)}. \label{eq:cl-avg}
\end{eqnarray}
In Fig.~\ref{fig:power_spectra}, we plot the averaged \hi signal power spectrum at the frequency of $1255\,{\rm MHz}$, and the sample variance of the power spectrum from simulation. As one can see, due to the cosmic variance on large angular scales, the intrinsic dispersion at low-$\ell$ is considerably larger than high-$\ell$.

\subsection{Noise}

\subsubsection{Thermal noise}
\label{thermal_noise_subsec}
We model the thermal noise (radiometer noise) as a white noise caused by the telescope system. Thermal noise is related to the telescope system noise and band width. For each pixel, the thermal noise can be approximated as Gaussian
white noise with {\it rms} amplitude~\citep{Wilson_2009} as
 \begin{equation}\label{sigma_pix}
  \sigma_{\rm pix} = \frac{T_{\rm sys}}{\sqrt{\Delta \nu t_{\rm pix}}},
 \end{equation}
where $T_{\rm sys}$ is the system temperature, $t_{\rm pix}$ is the integration time for each pixel and $\triangle \nu$ is the frequency bandwidth (frequency resolution).

A pixel size is given by its full-width at half-maximum (FWHM)
\begin{eqnarray}
 \theta_{\rm FWHM} &=& \frac{1.22 \lambda_{\nu}}{D} \nonumber \\
&=&  1.22 \frac{21 \ {\rm cm} (1+z)}{300 \ {\rm m}} \nonumber \\
 & = & 2.94\left(\frac{\nu_{10}}{\nu}\right) \ {\rm arcmin},
\end{eqnarray}
where $\lambda_{\nu}=21\,{\rm cm}(1+z)$ is the wavelength at the receiver, $D$ is the telescope's illuminated aperture.

The integration time for each pixel, $t_{\rm pix}$, for the assumed total observational time $t_{\rm obs} = 1\,{\rm year}$ and survey area, $\Omega_{\rm sur} = 20,000 \ {\rm deg}^{2}$ is
\begin{equation}
\begin{split}
 t_{\rm pix} &\quad = n_{\rm f}t_{\rm obs} \frac{\Omega_{\rm pix}}{\Omega_{\rm sur}}\\
  &\quad = 71.98\times \Bigg[  \frac{1420 \ {\rm MHz}}{\nu} \Bigg]^{2} \ {\rm s},
\end{split}
\end{equation}
where $\Omega_{\rm pix} \equiv \theta_{\rm FWHM}^{2}$ is the beam area. We can therefore calculate the {\it rms} by substituting $t_{\rm pix}$ values into Eq.~(\ref{sigma_pix}). We will use the Python Healpy to generate the noise maps at different frequencies, taking into account a particular number of sky pixels ($N_{\rm side} = 256$).

\subsubsection{1/f Noise}
\label{f_noise_subsec}
\begin{figure}
\includegraphics[width=8cm]{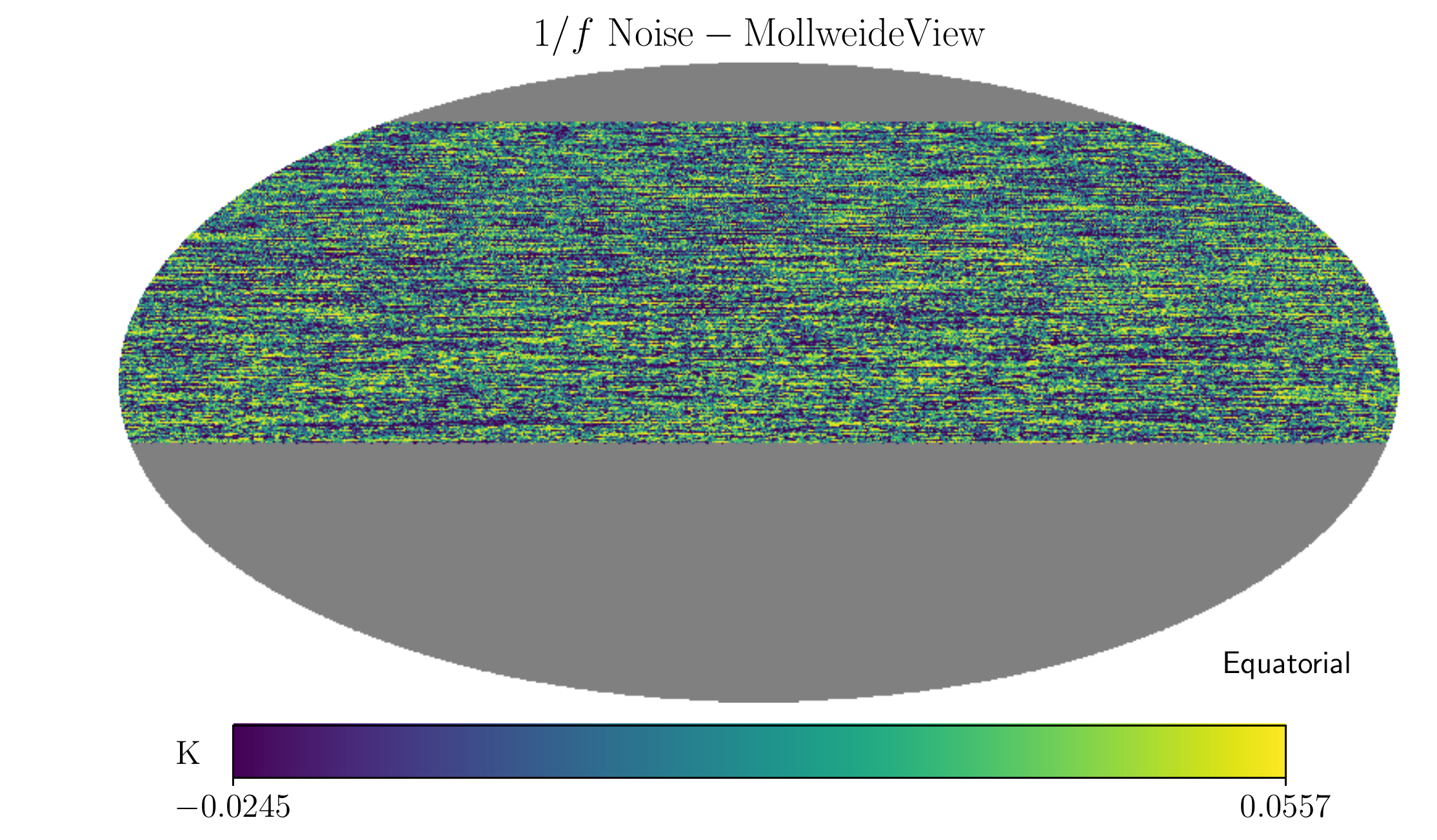}
\caption{FAST mollweide projection of the $1/f$ noise at frequency, $\nu = 1,250\,{\rm MHz}$, for parameters $\beta = 0.25$, $\alpha = 1.0$ and a knee frequency $f_{\rm k} = 1.0\,{\rm Hz}$.}
\label{1_over_f_mollview_map}
\end{figure}

$1/f$ noise is an instrumental effect different from the thermal noise. This is the correlated noise across frequency bands, mostly affects radio receiver
systems, revealing itself as small gain fluctuations~\citep{Harper_2018}. Binning $1/f$ noise on the sky map can result in apparent spatial fluctuations that resemble large-scale structure signal, which is a potential confusion effect (Fig.~\ref{1_over_f_mollview_map}).

The power spectral density (PSD) for the thermal and $1/f$ noises which takes into account both the temporal and spectroscopic correlations is
\begin{eqnarray}
\label{1_over_f_eqn}
 {\rm PSD}(f) = \frac{T_{\rm sys}^{2}}{\Delta \nu }\left[1 + C(\beta, N_{\nu})\left(\frac{f_{\rm k}}{f}  \right)^{\alpha} \left(\frac{1}{\omega \Delta B} \right)^{\frac{1-\beta}{\beta}} \right].
\end{eqnarray}
Here, $T_{\rm sys}$ is the system temperature, $\Delta \nu$ is the frequency channel bandwidth, $f_{\rm k}$ is the knee
frequency, and $\alpha$ is the spectral index of temporal fluctuations. The unity term in Eq.~(\ref{1_over_f_eqn}) describes the contribution by the thermal noise, and the
reciprocal power-law $\left(f_{\rm k}/f \right)^{\alpha}$ describes the $1/f$ noise. When $\alpha >0$, it
implies that the power gained is proportional to time-scale fluctuations.
Furthermore, different values of the spectral index $\alpha$ characterize several variations in the names of $1/f$ noise, in particular, pink noise, brown noise, and
red noise are respectively characterized by $\alpha = 1$, $\alpha = 2$, and generally for any value of $\alpha > 0$.

Moreover, $\omega$ is the inverse spectroscopic frequency wavenumber, $\Delta B$ is the total receiver bandwidth, $\beta$ is the PSD spectral index parametrization parameter, and $C(\beta, N_{\nu})$ is a constant, described in detail in~\cite{Harper_2018}.
$0 < \beta < 1$ is the limit for which the spectral index
of the frequency correlation is defined, where small values of $\beta$ indicates high correlations, such that $\beta = 0$ implies
identical $1/f$ across different frequency channels, and $\beta = 1$ would describe independent $1/f$ noise in every frequency channel.
For a complete detailed account of $1/f$ noise including its modeling we refer the interested reader to~\cite{Harper_2018}.

Generation of $1/f$ noise was carried out by using the end to end simulations (simulations assume that $1/f$ noise fluctuations have some Gaussian properties)
described in~\cite{Harper_2018} as would be required by most of the time-dependent systematics. The approach of modelling sky signal by~\cite{Harper_2018}, however, intends to account
for the maximum impact of $1/f$ noise for a particular telescope model on the recovery of the \hi signal spectrum using component separation techniques.

\subsection{Foreground templates}

It is a common understanding that the biggest challenge of using the 21-cm IM technique is to develop a computationally effective strategy to remove the foreground contaminants.
The foreground contaminants include, but not limited to, Galactic synchrotron emission, emitted by electrons spiralling in Galactic magnetic field~\citep{Pacholczyk_1970, Banday_1990, Banday_1991};
radiation from the background of extragalactic point/radio sources (unresolved foreground) that includes a mixture of radio galaxies, quasars and other objects; and free-free radio emission produced by free electrons that encounter ions and scattering off them without being captured. Among these foregrounds, Galactic synchrotron emission is the most notable and overshadows the \hi  signal of interest here by several orders of magnitude~\citep{Bigot-Sazy_2015}. There is however thermal/white noise and an instrumental $1/f$ noise (see subsections \ref{thermal_noise_subsec}, \ref{f_noise_subsec}), radio frequency interference (RFI), time-variable
noise introduced during propagation of the signal through the atmosphere which additionally contributes to the $1/f$ noise of the instrument, and atmospheric effects caused by absorption or scattering of signals and fluctuations arising from the turbulence in the emission of the water vapour~\citep{Bigot-Sazy_2015}. We explain some of these contaminants in the subsequent subsections.

\subsubsection{Galactic synchrotron emission}

Galactic synchrotron emission varies across the large scale of the sky, characterized by the quadrupole features and additional signal in the Galactic plane. It also varies as a function of frequency. A template for Galactic synchrotron sky emission can be generated by extrapolating/interpolating at appropriate frequencies the all-sky $408$ MHz continuum Haslam map. To date, the Haslam map and the reprocessed all-sky $408$ MHz map are publicly available\footnote{\url{http://www.jb.man.ac.uk/research/cosmos/haslam_map/}}~\citep{Haslam_1982}. The reprocessed and improved $408$ MHz all-sky map continues to offer the best approximation and characterization of the diffuse Galactic synchrotron emission.


Following the framework presented in~\cite{Shaw_2014}, the global sky map~\citep{de_Oliveira-Costa_2008} generated by compiling maps from $10$ MHz to $94$ GHz is used to generate sky temperature maps at frequencies $400$ MHz and $1420$ MHz.
These maps are then used for the calculation of an effective spectral index at each sky location, $\hat{\bf n}$, estimated as
\begin{equation}
 \alpha(\hat{\bf n}) = \frac{{\rm log}T_{\rm 1420}(\hat{\bf n}) - {\rm log}T_{\rm 400}(\hat{\bf n})}{{\rm log}1420 - {\rm log}400}.
\end{equation}
The spectral index is used in combination with the $408$ MHz map~\citep{Haslam_1982} to extrapolate sky temperature maps at different frequencies using the power
law~\citep{de_Oliveira-Costa_2008, Bigot-Sazy_2015, GNILC}:
\begin{equation}\label{plaw_extrap}
T(\hat{\bf n}, \nu) = T_{\rm 408}(\hat{\bf n})\left( \frac{\nu}{408 \ {\rm MHz}} \right)^{\alpha(\hat{\bf n})}.
\end{equation}
The Galactic synchrotron model simulated by~\cite{Shaw_2015} is suitably calibrated for both Galactic plane and low frequencies, and the resulted model has been transitioned from
low to higher frequencies as described in~\cite{Shaw_2014} to make the angular power law applicable for \hi IM simulations.


Previously, many cosmologists simply extrapolated the $408$-MHz Haslam maps to lower/higher frequencies and ignored any predicted spectral variations across the sky. Although Galactic synchrotron emission is expected to dominate at low frequencies, it has been observed that its dependence on frequency is not a perfect power-law, instead, the
slope of the Galactic synchrotron emission progressively steepens with an increase in frequency, at the same time other Galactic contaminants such as free-free and dust emissions noticeably start to trickle in.
Due to the fact that this power-law extrapolation (Eq.~(\ref{plaw_extrap})) does not take into account any spectral variations, the resultant maps lack the small scale angular
fluctuations because of the limited resolution of
the Haslam map template used~\citep{Shaw_2014}. These missing expected real sky components have been included for realistic foreground model tests as described in~\cite{Shaw_2014, Shaw_2015}.

In light of this observation, we use the Cosmology in the Radio Band (CORA) code developed by~\cite{Shaw_2015} which takes into account the radio emission spectral variations and small-scale angular fluctuations to simulate Galactic synchrotron emission templates (point sources and 21-cm as well) for our foreground removal with PCA.

\subsubsection{Extragalactic point sources}
We summarize the analysis by~\cite{Shaw_2014, Shaw_2015}, where the extragalactic point sources are assumed to be an isotropic field modelled as the power law in both frequency and multipole moment, $\ell$,
\begin{eqnarray}\label{syn_ext_ps_eq}
 C_{\ell}(\nu, \nu') = A\left( \frac{100}{\ell}\right)^{\alpha}\left(\frac{\nu \nu'}{\nu_{0}^{2}} \right)^{-\beta}{\rm exp}\left[-\frac{1}{2\xi_{\ell}^{2}}{\rm ln}^{2}\left(\frac{\nu}{\nu'} \right)  \right],
\end{eqnarray}
which was originally described by~\cite{Santos_2005} and applied to low frequencies during the Epoch of Reionization, and later modified in~\cite{Shaw_2014, Shaw_2015} to suit high frequencies and the full-sky
intensity mapping regime.
Here $C_{\ell}$ is the angular power spectrum, and $\nu$, $\nu'$ represent
two different frequency bands with $\nu_{0}$ being a pivot frequency.

The approach uses simulated maps of the point sources which are composed of two different populations. The first population is constructed directly
following the point sources distribution by~\cite{Di_Matteo_2002} and forms a population of bright and isolated point sources with a flux $S > 0.1 \ {\rm Jy}$ at $151$ MHz,
and the second one is a continuum/background of dimmer unresolved points sources (whose flux $S < 0.1 \ {\rm Jy}$), simulated by drawing random realization (Gaussian random field)
from Eq.~(\ref{syn_ext_ps_eq}) by adopting parametrization from \cite{Shaw_2014}, where $A = 3.55 \times 10^{-4}\,{\rm K}^{2}$, $\alpha = 2.1$, and the spectral index $\beta = 1.1$.
Here, the parameter $\xi_{\ell}$ measures the foreground frequency coherence/correlation. Two limits of this parameter
$\xi \to 0$ and $\xi \to \infty$ represents the limits of complete foreground-frequency incoherence and perfect foreground-frequency coherence~\citep{Tegmark_2000, Santos_2005}.
This foreground-frequency correlation length parameter $\xi$ can be determined in terms of the spectral index $\beta$ as described in detail in~\citet{Tegmark_1998},~\citet{Tegmark_2000} and~\cite{Santos_2005}. Such treatment is important because it takes into account possible changes of the foregrounds with the observed direction/position on the sky, and the relative power ratio between various sky components
which may also vary with the angular scales~\citep{GNILC}.

In the former population \citep{Di_Matteo_2002}, sources are randomly distributed over the sky, where a pure power-law emission is assumed to model each source
with a randomized spectral index~\citep{Shaw_2015}.

In practice, the brightest radio sources ($S>10 \ {\rm Jy}$) above the threshold flux are usually subtracted or masked. In order for the Di Matteo model~\citep{Di_Matteo_2002} point sources distribution to be useful in a range of higher frequencies, and also to be able to
adjust the maximum flux of sources (that were not subtracted) from $0.1 \ {\rm mJy}$ to $0.1 \ {\rm Jy}$, the pivot frequency is changed from $150$ MHz to Haslam $408$ MHz frequency, and the amplitude $A$ rescaled.

\subsubsection{Free-free emission}
Free-free emission arises due to the scattering between ions and free electrons in the ionized medium. The term free-free follows from the nature of the emission, in which electrons are free before they encounter ions
and thereafter scatter off ions and remain free again~\citep{Rybicki1979, GNILC}. These electrons seen in radio frequencies, are originated
from warm ionized gas whose temperature $T_{\rm e} \simeq 10^{4}$\,K~\citep{GNILC}.

According to~\cite{Dickinson_2003} and~\cite{GNILC}, in an electrically charged medium of ions and electrons, free-free emission is scaled by frequency as
\begin{eqnarray}
 T_{\rm ff} \approx 90 \ {\rm mK}\Bigg(\frac{T_{\rm e}}{{\rm K}}\Bigg)^{-0.35}\Bigg(\frac{\nu}{\rm GHz} \Bigg)^{\beta}\Bigg(\frac{\rm EM}{{\rm cm}^{-6}{\rm pc}}\Bigg),
\end{eqnarray}
where $\nu$ is frequency, and ${\rm EM} = \int n_{\rm e}^{2}{\rm d}\ell$ is called emission measure, interpreted as the integral of the electron density squared along the line of sight~\citep{GNILC}, and
$\beta \sim -2.1$ is the spectral index. We estimate the emission measure (EM) and generate the Galactic free-free temperature maps by using the base Wisconsin H-Alpha Mapper (WHAM) survey maps,
where we have considered an electron temperature, $T_{\rm e} = 7000 \ {\rm K}$.

\begin{figure*}
\includegraphics[width=16cm]{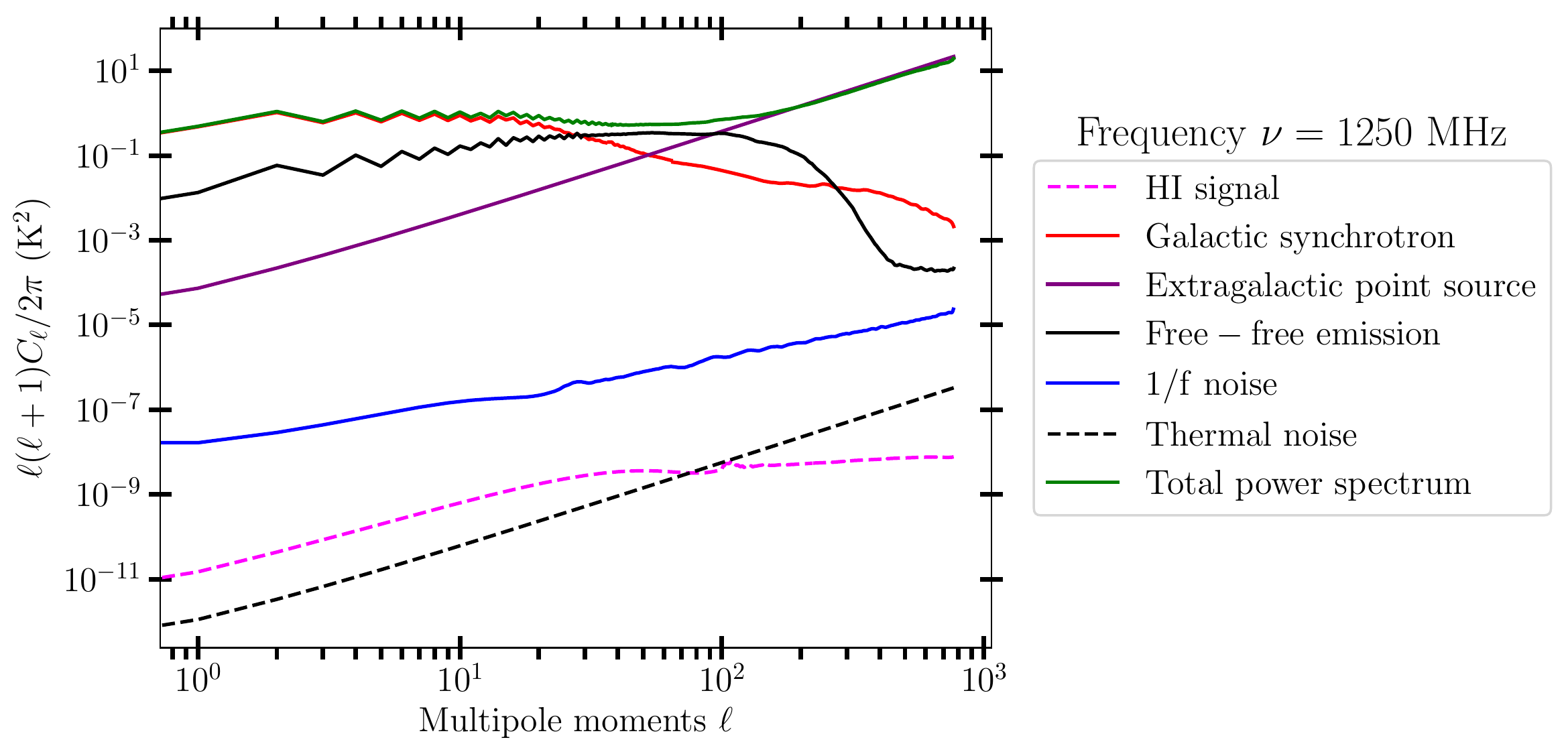}
\caption{FAST full-sky approximation power spectra for galactic synchrotron, extragalactic point sources, $1/f$ noise, free-free emission and \hi signal
simulated at the FAST bandwidth mid-range frequency, $\nu = 1.25$ GHz.}
\label{power_spectra}
\end{figure*}

We present in Fig.~\ref{power_spectra} the power spectra for some of these notable contaminants; as we already discussed, these components are simulated by assuming a full-sky approximation.

\subsection{Sky Area}
We use the Equatorial coordinate system (right ascension (RA) and declination (DEC)) to specify points and direction on the celestial sphere in this coordinate system.

We consider the FAST maximum declination range $[-14^{\circ}, 65^{\circ}]$, which means we utilize the full sky region that is potentially surveyable by FAST telescope, as shown in Fig.~\ref{contaminated_map}.

\section{Principal Component Analysis}
\label{PCA_algorithm}
To illustrate the foreground subtraction algorithm and capture various effects for a wide range of frequency variation, we will perform foreground subtraction over the full FAST
frequency bandwidth, $1,050 - 1,450$ MHz,
but report results 
for the middle ($1250$ MHz) frequency (or equivalently an average frequency in the band).
Principal component analysis (PCA) is a simple non-parametric method for extracting useful information from a high-dimensional dataset.
The method is a multivariate statistical procedure that finds the direction of maximum variance by orthogonally transforming
a possibly correlated high-dimensional dataset of observations into a low-dimensional linearly-uncorrelated subspaces called principal components.

This process involves compressing a lot of data by projecting it into a smaller dimensional subspace while retaining the essence of the original data. In our case, we will apply PCA to transform noisy data into a
subspace that consists of two measurement object patterns, one composed of dominant components (the foregrounds) and the other one consisting of the complimentary component, i.e. the \hi signal. The scientific information that will be collected by the radio telescopes is expected to be highly contaminated, and thus devising means to subtract foregrounds and noises from \hi signal is essentially very important at this radio astronomy developmental stage.

We encode our total emission sky dataset as a matrix $X$ with dimensions $N_{\nu} \times N_{\rm p}$, where $N_{\nu}$ is the number of frequency channels, and $N_{\rm p}$ is the number of pixels of the temperature
fluctuation map. We can think of this matrix as composed of $N_{\nu}$ samples and $N_{\rm p}$ pixelized temperature measurements of the brightness temperature, $T(\nu, \hat{n}_{\rm p})$, corresponding to the frequency $\nu$, and along
the direction of the line of sight $\hat{n}_{\rm p}$~\citep{Bigot-Sazy_2015}. To distinguish the frequency with spatial indices, we use Greek symbols for frequency index, and Latin symbol to denote spatial index.

We can then compute the covariance matrix as
\begin{align}\label{cov_mat2}
 C = \frac{1}{N_{\rm p}}\left( \left(X - \mu \right) \left(X - \mu  \right)^{T}\right),
\end{align}
where $\mu$ is the population mean. Scaling observations by $N_{\rm p} - 1$ is usually considered as a correction for the bias introduced when the sample mean is used instead of the population mean.

Next, we can normalize the covariance matrix (Eq.~(\ref{cov_mat2})) by calculating the entries of the correlation matrix between each pair of
frequency channels
\begin{align}\label{corr_mat}
 r_{\alpha \beta} = \frac{\sigma_{\alpha \beta}}{\sqrt{\sigma_{\alpha \alpha}}\sqrt{\sigma_{\beta \beta}}},
\end{align}
where $\sigma_{\alpha \alpha} \equiv \sigma_{\alpha}^{2}$ is the variance (covariance of a variable with itself). $r_{\alpha \beta}$ are entries of the dimensionless correlation matrix, $R_{\alpha \beta}$, such that $-1\le r_{\alpha \beta} \le 1$,
$r_{\alpha \alpha} = 1$, and can be interpreted as correlation coefficients between frequency pairs. The quantities $\sigma_{\alpha}$ characterize the {\it rms} fluctuations at each frequency.

The eigenvectors of a correlation matrix (Eq.~(\ref{corr_mat})) forms the basis for the principal component analysis. The vectors
in the reduced subspace determine the new axis directions, and the magnitudes of their corresponding eigenvalues describe the variance of the data of the resulting
subspace axis. Therefore, PCA requires that we perform the eigendecomposition on the correlation matrix (Eq.~(\ref{corr_mat})) or the covariance matrix (Eq.~(\ref{cov_mat2})). Similarly, PCA can be carried out after
performing SVD (singular vector/value decomposition) on the correlation/covariance matrix for the sake of computational efficiency and numerical robustness.

\begin{figure*}
\includegraphics[width=8.8cm]{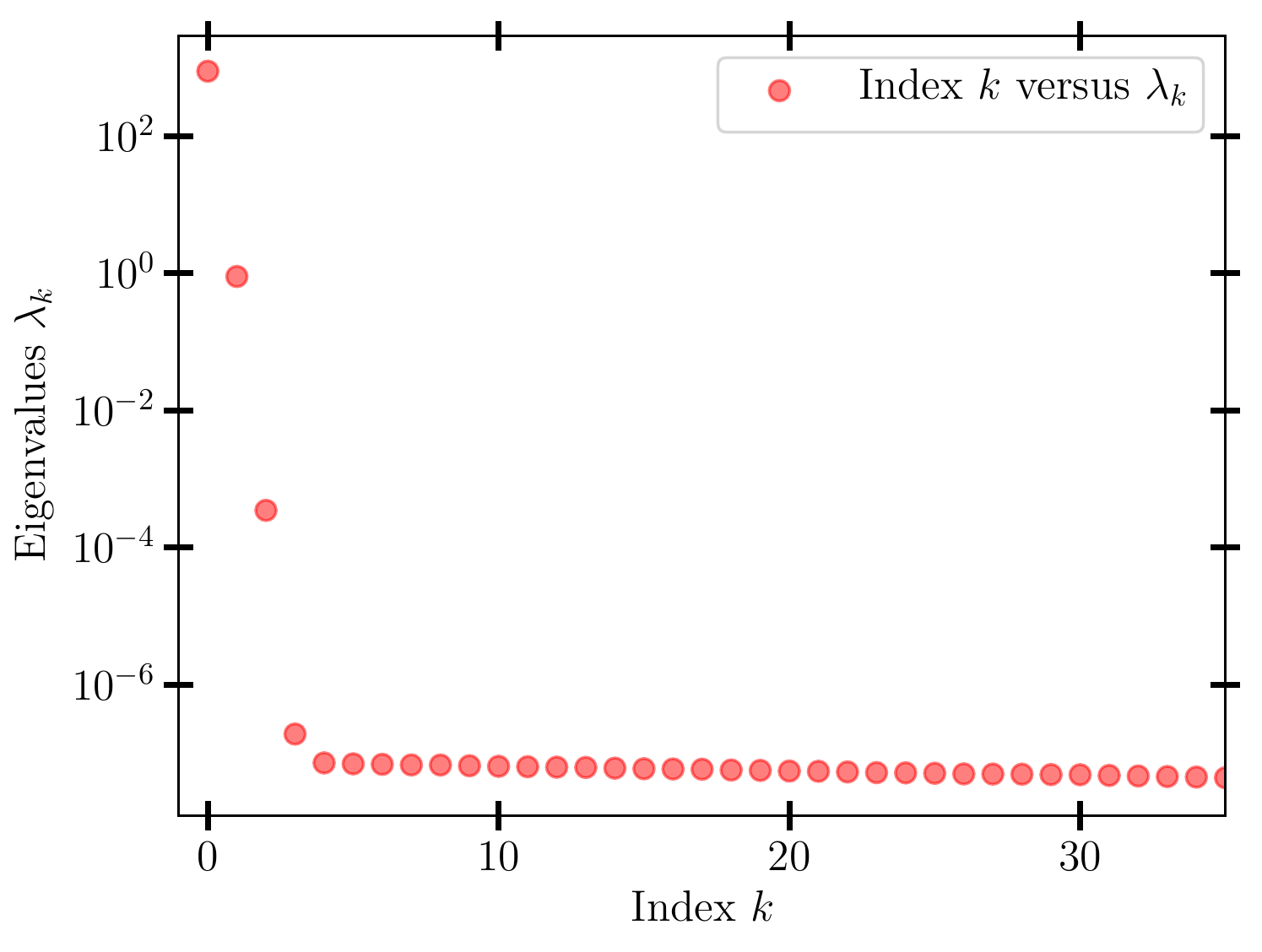}
\includegraphics[width=8.8cm]{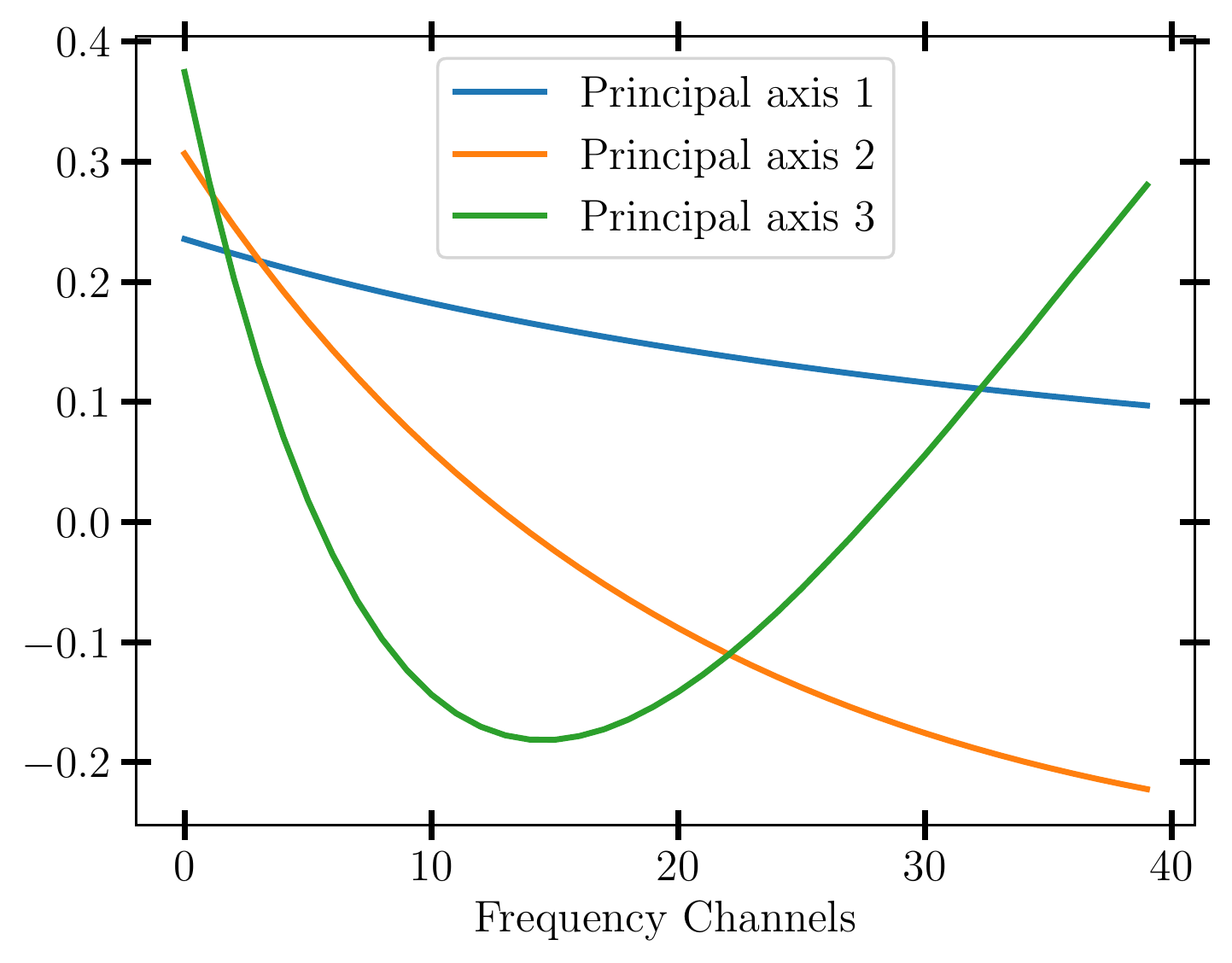}\\
\caption{{\it Left}--The eigenvalues profile corresponding to the $N_{\nu} \times N_{\nu}$ matrix of eigenvectors, used for PCA with FAST. {\it Right}--The principal axes corresponding to the first three eigenvalues of this matrix.
Because the foreground dominates the sky map, it is represented by the largest principal components; in this case, the first four principal components contain more than $99\%$ of the total foreground information.}
\label{eigenvalues_evolution}
\end{figure*}

However, the magnitude of the eigenvalues can give us clues on which eigenvectors (principal axes) correspond to the dominant
foregrounds. This can easily be seen by ranking the eigenvectors in the decreasing order of their corresponding eigenvalues (see the left panel of Fig.~\ref{eigenvalues_evolution}). Usually, the first
few principal components can be attributed to the dominant
variance (information) -- the foregrounds.

As previously stated, the eigendecomposition can be carried on either the covariance matrix or correlation matrix, depending on which one is preferred for PCA.
Here, we illustrate these cases by proceeding with the diagonalization of the covariance matrix,
\begin{align}\label{eigen_decompose}
C = W^{T}\Lambda W,
\end{align}
where $WW^{T} = W^{T}W = I$, implying that $W$ is orthogonal matrix, whose columns are the principal axes/directions (eigenvectors) (see the right panel of Fig.~\ref{eigenvalues_evolution}), and the diagonal matrix of the corresponding eigenvalues is given by
$\Lambda_{\alpha \beta} = \delta_{\alpha \beta}\lambda_{\alpha}$. After identifying the principal axes, we next compose an $N_{\nu} \times k$ ($k < N_{\nu}$) matrix, $W'$, called the projection matrix,
a matrix whose columns are made up of the first $k$ columns of $W$, that will form the dominant principal components components when the data matrix $X$ is projected onto them.

Finally, we project our data matrix $X$ onto a new subspace by using the projection matrix $W'$ via the equations
\begin{align}
 U = {W'}^{T}\cdot X,
\end{align}
\begin{align}
 V = W'\cdot U,
\end{align}
and recover the \hi signal as
\begin{align}
 S_{\rm HI} = X -  V,
\end{align}
where $V$ is the map of the reconstructed foreground. Lastly, we project the patch of the cleaned \hi signal pixels into the correct position in the sky map.

More often, singular value decomposition (SVD) is favored, applied just in the same way as PCA to a real or complex rectangular matrix, but with more computational power. SVD decomposes a data matrix $X$ in the form
\begin{align}\label{complex_v}
 X = W^{*}\Sigma R,
\end{align}
where $W^{*}$ and $R$ are unitary matrices, i.e. $WW^{*} = I$, $RR^{*} = I$, and they are respectively, called left and right singular vectors; $\Sigma$ is a rectangular diagonal matrix of singular values.
In this form (Eq.~(\ref{complex_v})), $W$ is generally complex-valued matrix and $*$ denotes conjugate transpose. Since we are dealing with a real data matrix $X$, the resulting unitary matrices will be real, and thus
(Eq.~(\ref{complex_v})) takes the form
\begin{align}\label{real_v}
 X = W^{T}\Sigma R.
\end{align}
With SVD, we bypass calculations of the covariance matrix, by looking for something equivalent to it in a computationally efficient way. This requires application of SVD on the covariance matrix $C$ as
\begin{align}
 \begin{split}
  C  &\quad = \frac{XX^{\rm T}}{N_{\rm p} - 1}\\
  &\quad = \frac{(W^{T}\Sigma R)(W^{\rm T}\Sigma R)^{\rm T}}{N_{\rm p} - 1}\\
  &\quad = \frac{W^{\rm T}\Sigma^{2}W}{N_{\rm p} - 1}\\
  &\quad = \frac{W^{-1}\Sigma^{2}W}{N_{\rm p}-1},\\
 \end{split}
\end{align}
where the last equality follows from the fact that $W$ is unitary.

We see that the result takes the form of the eigendecomposition (Eq.~(\ref{eigen_decompose})), and we can easily notice the relationship between the eigenvalues
$\Lambda$ and the singular values $\Sigma$, where we establish that
\begin{align}
 \Lambda = \frac{\Sigma^{2}}{N_{\rm p} - 1},
\end{align}
implying $\lambda_{\alpha} = \sigma_{\alpha}^{2}/N_{p}-1$ for $\lambda_{\alpha} \in \Lambda$ and $\sigma_{\alpha}\in \Sigma$. Principal components are given by $WX = WW^{\rm T}\Sigma R = \Sigma R$, and
singular values can be arranged in decreasing order $\sigma_{1}>\sigma_{2}> \sigma_{3} \hdots$, such that the first column of the principal components $\Sigma R$
corresponds to the first singular value, and so on. Thereafter, PCA can be performed under this new transformation. In this work, we will favor PCA over
covariance matrix (Eq.~(\ref{cov_mat2})) for very rapid convergence of
the \hi signal recovery process, but this is only true if the foreground is not too complicated as we shall see in the subsequent sections.

We describe the performance of PCA results in Section~\ref{HI_recovery}, obtained by applying the algorithm to subtract foregrounds for the FAST telescope specifications.

\section{PCA Results}
\label{HI_recovery}

\begin{figure}
\includegraphics[width=8.8cm]{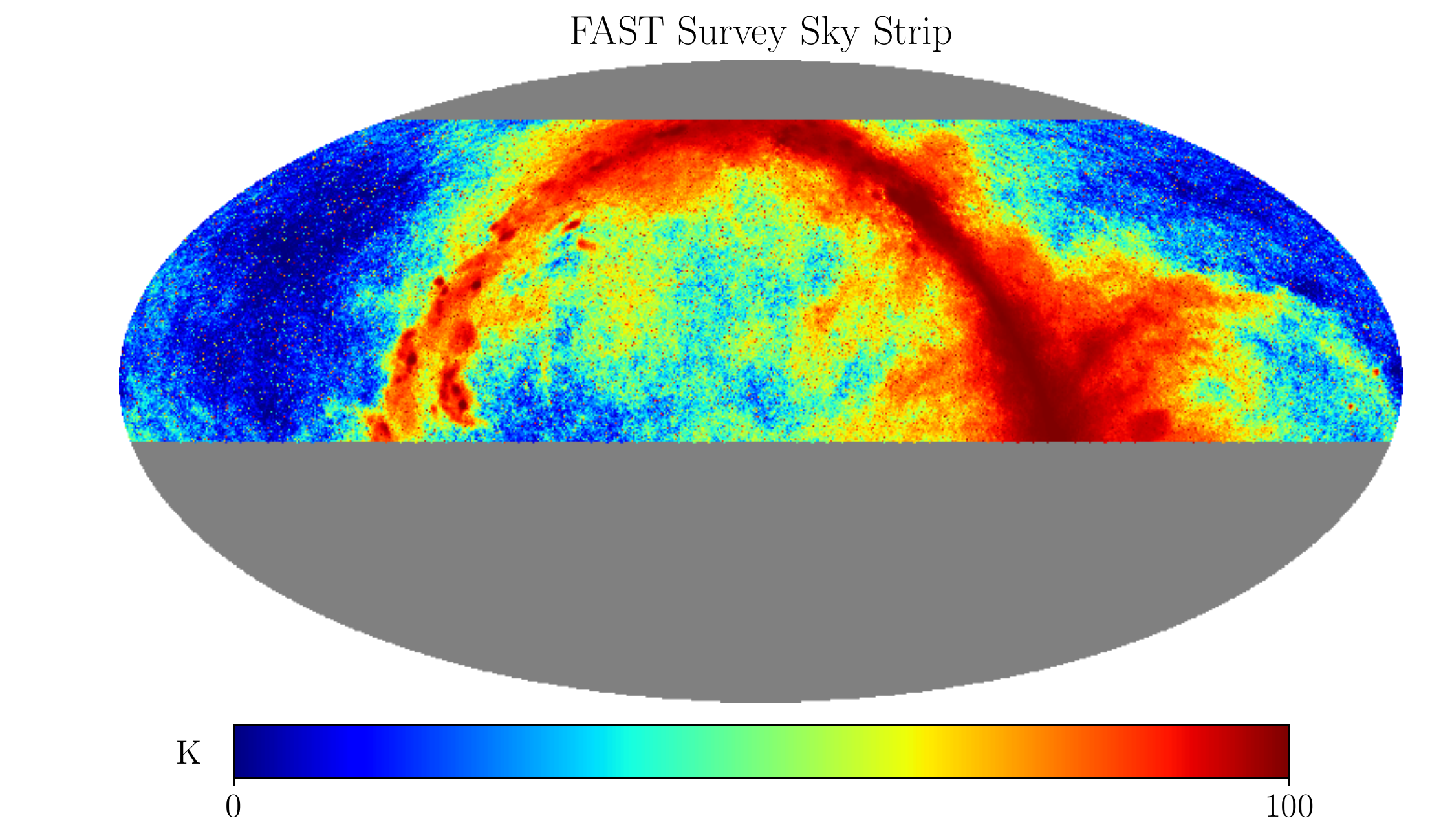}
\caption{Sky map containing anticipated components that are significant, within the FAST survey sky strip (Galactic synchrotron + extragalactic point sources + free-free emission + \hi signal + $1/f$ noise), simulated for FAST telescope at the frequency $1.25 \ {\rm GHz}$.}
\label{contaminated_map}
\end{figure}

\begin{figure}
\includegraphics[width=7cm]{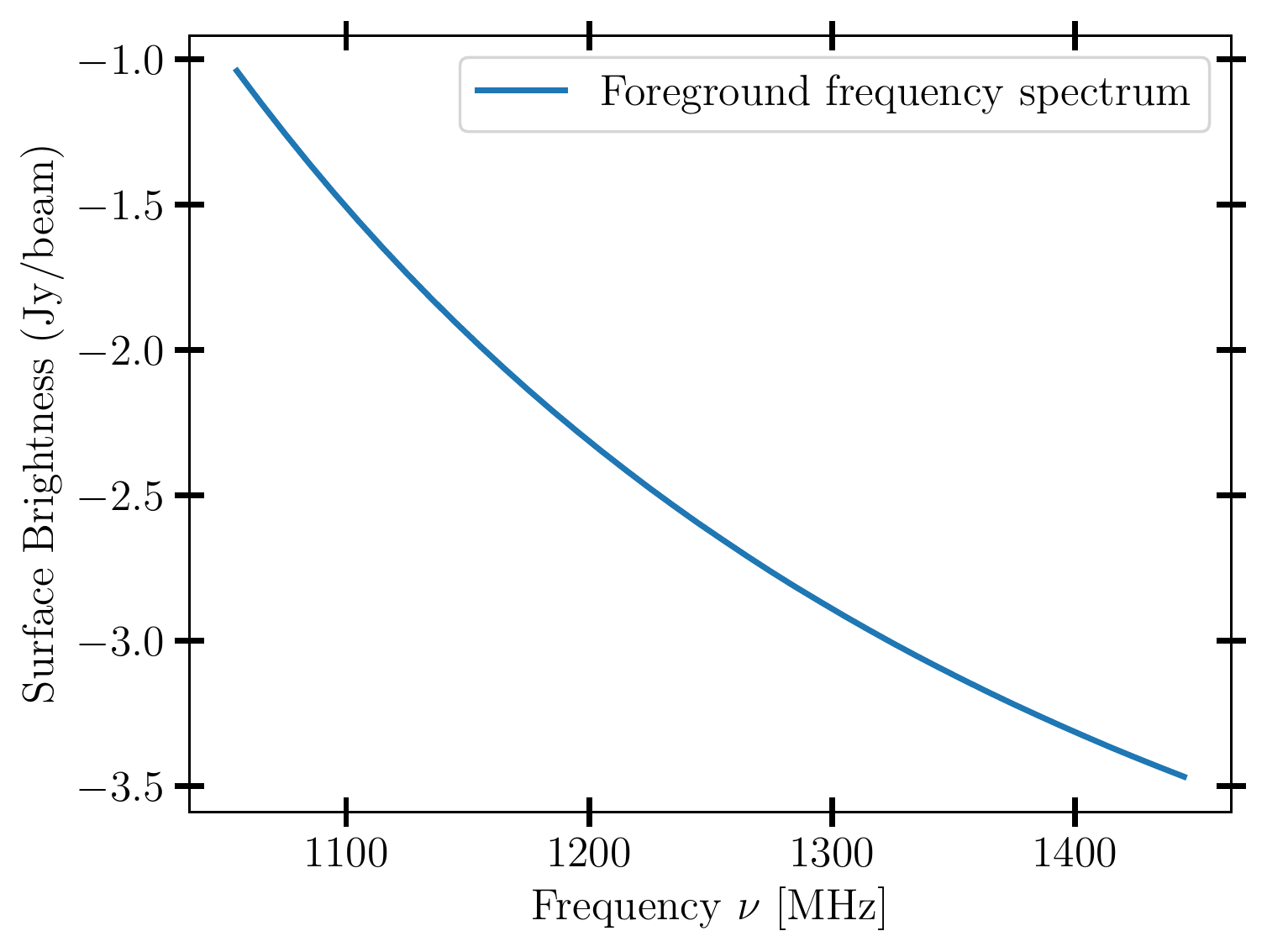}
\caption{FAST smooth foreground (Galactic synchrotron + extragalactic point sources + free-free emission) frequency spectrum, that is, the temperature flux at a given pixel. Foreground spectral smoothness feature greatly favours the process of decontaminating. High temperatures at lower frequencies are expected due to the Galactic foreground (mostly Galactic synchrotron) signal domination~\citep{Smoot_2017}.}
\label{smooth_foreground}
\end{figure}

\begin{figure*}
\centerline{\includegraphics[width=0.8\linewidth]{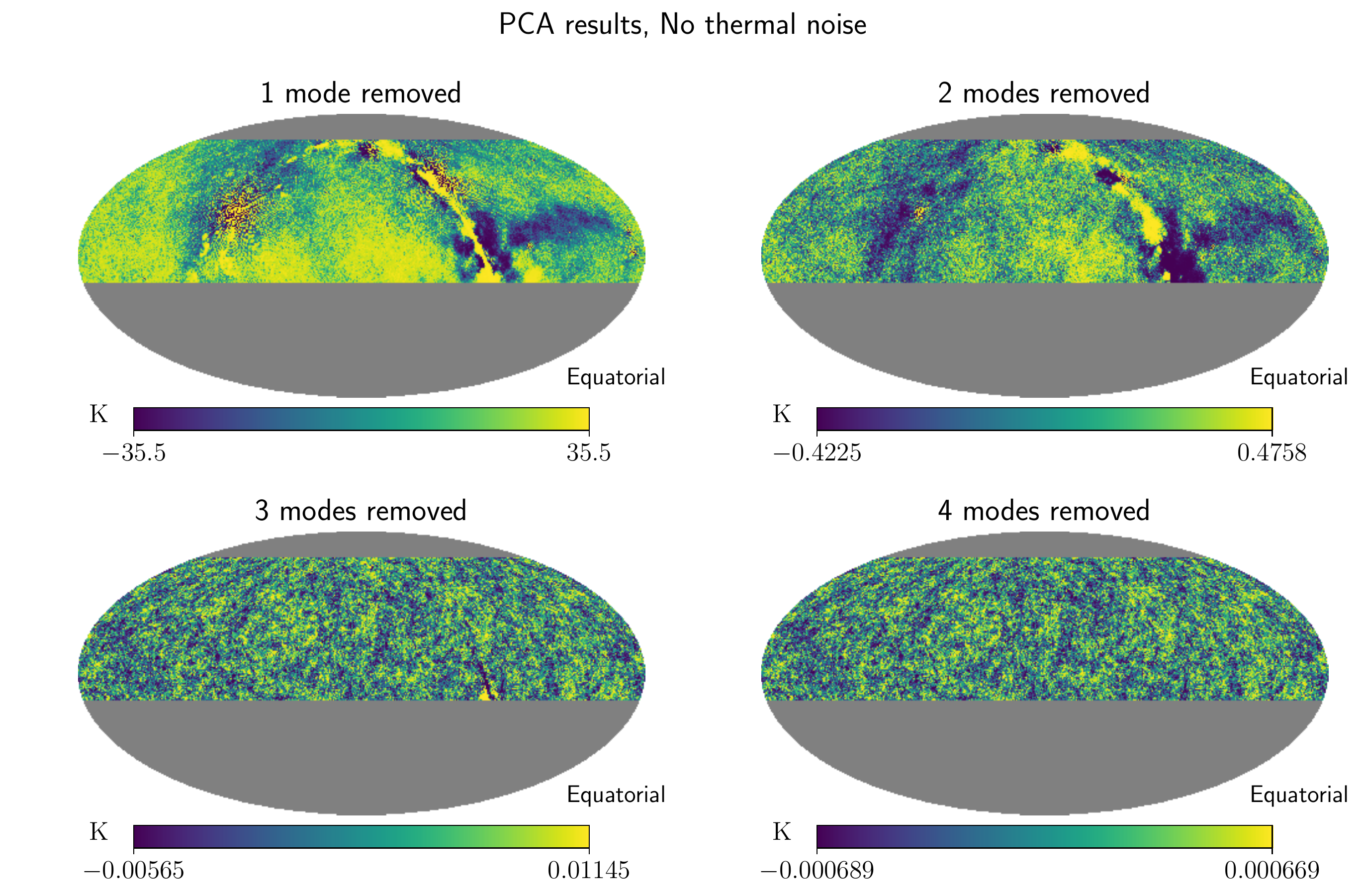}}
\caption{\hi sky map recovery when removing different numbers of PCA modes. The map is from a noise-free simulations at the frequency $1.25 \ {\rm GHz}$ and for the maximum multipole range $\ell = 768$.}
\label{fig:PCA_recovery_map}
\end{figure*}

\begin{figure}
\includegraphics[width=8.1cm]{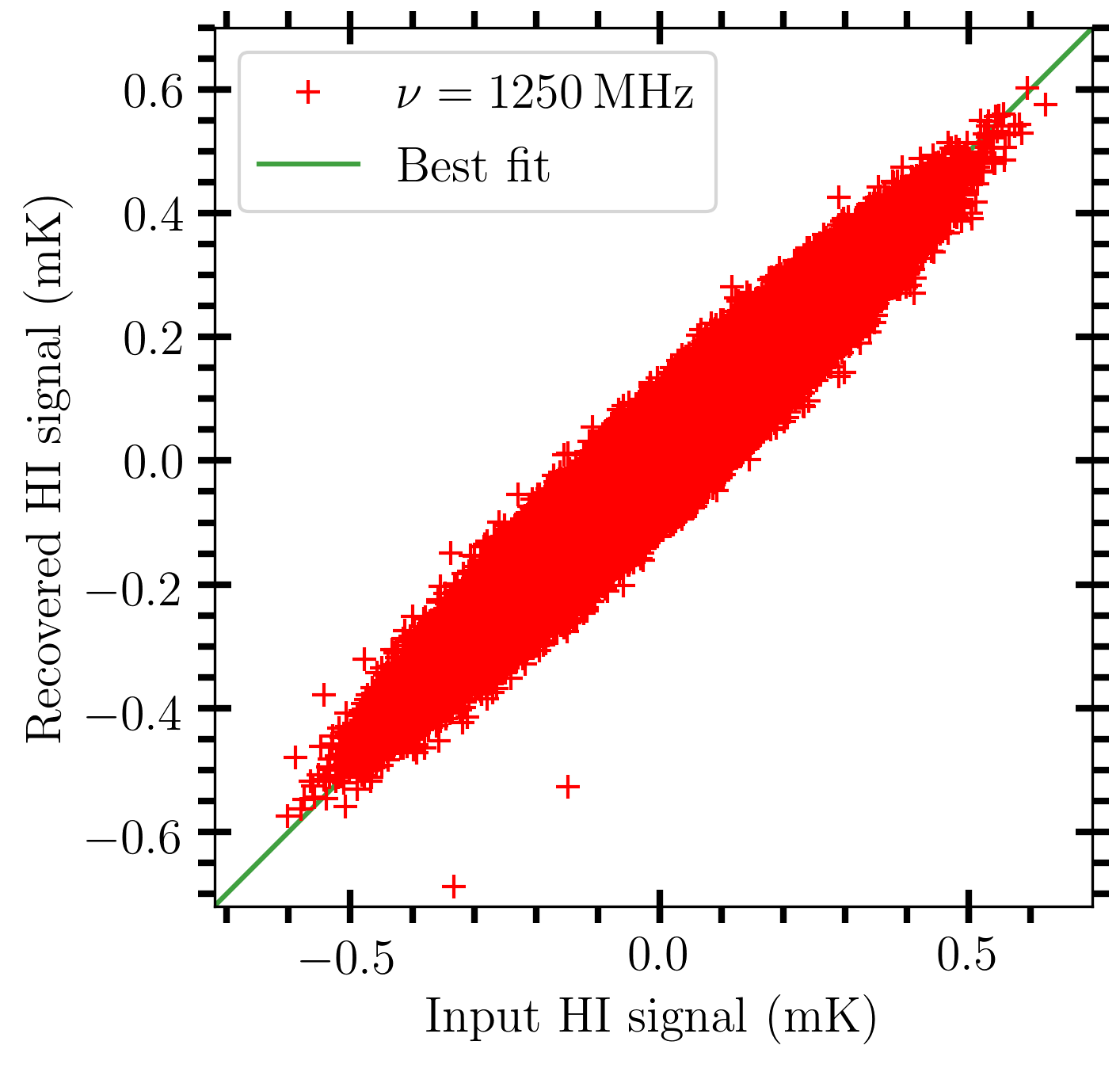}
\caption{Comparison between the input (simulated) \hi signal map, versus output (recovered) \hi signal map, showing the unbiased results of PCA analysis without the thermal noise inclusion in the foregrounds.
The standard dispersion between input and output signals is $\Delta T \equiv \sqrt{\sum_{i}\left(T^{\rm in}_{i}-T^{\rm out}_{i} \right)^{2}/N}=0.034\,{\rm mK}$,
indicating the robustness of PCA reconstruction.}
\label{Input_vs_Recovered_HI}
\end{figure}

\begin{figure*}
\centerline{\includegraphics[width=0.45\linewidth]{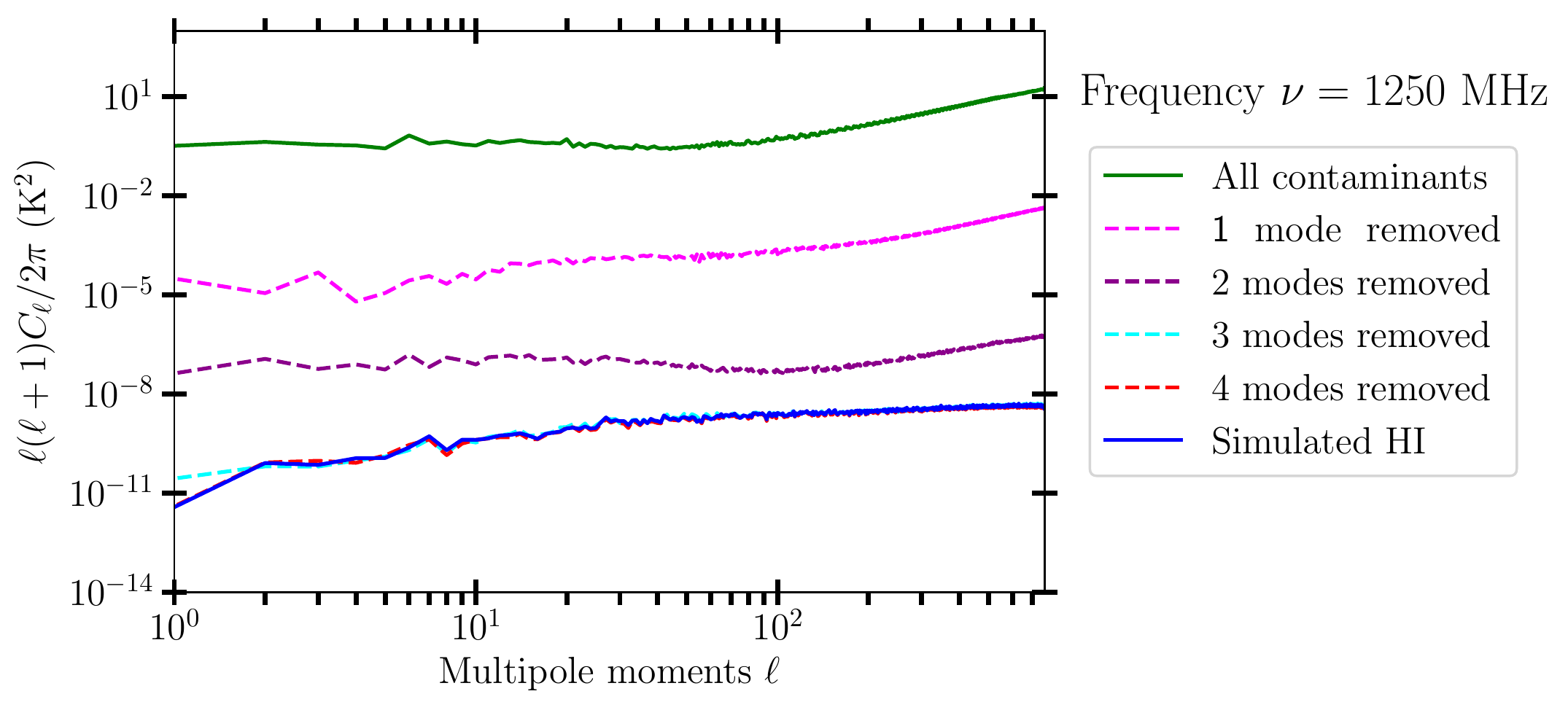}
\includegraphics[width=0.5\linewidth]{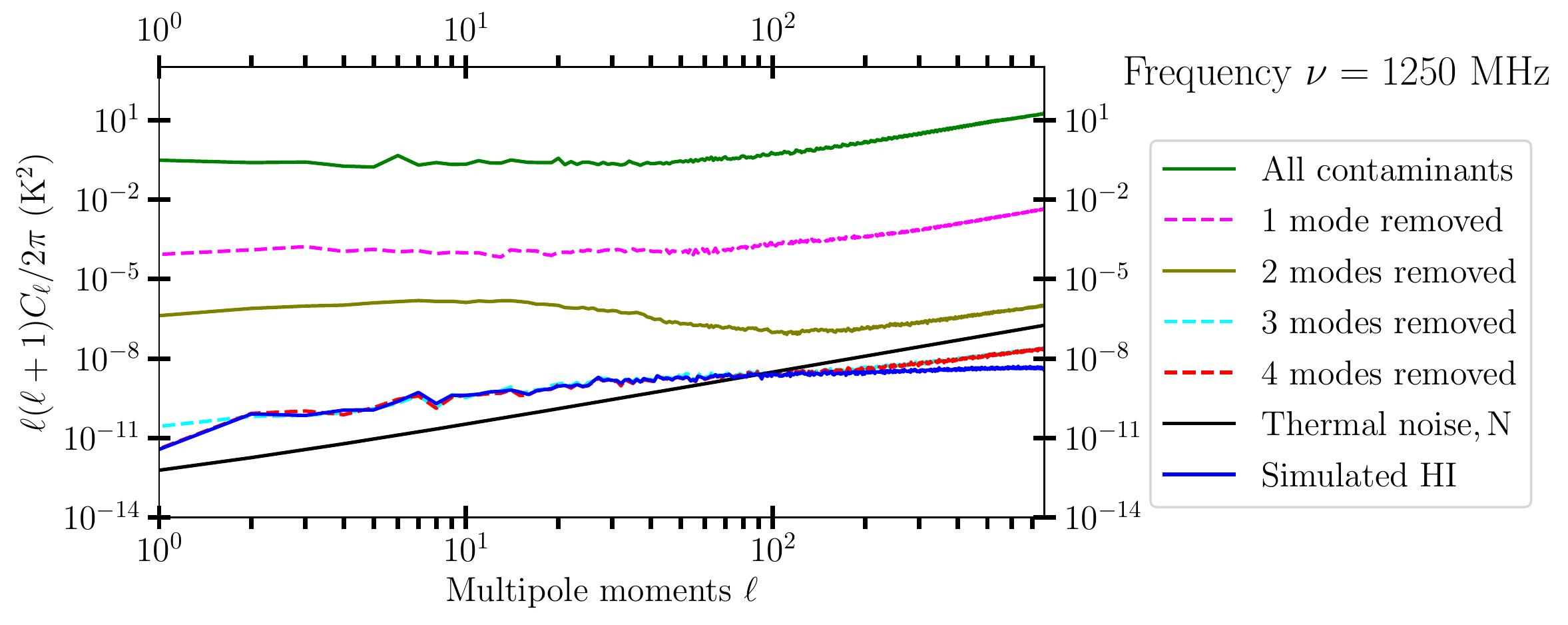}}
\centerline{\includegraphics[width=0.55\linewidth]{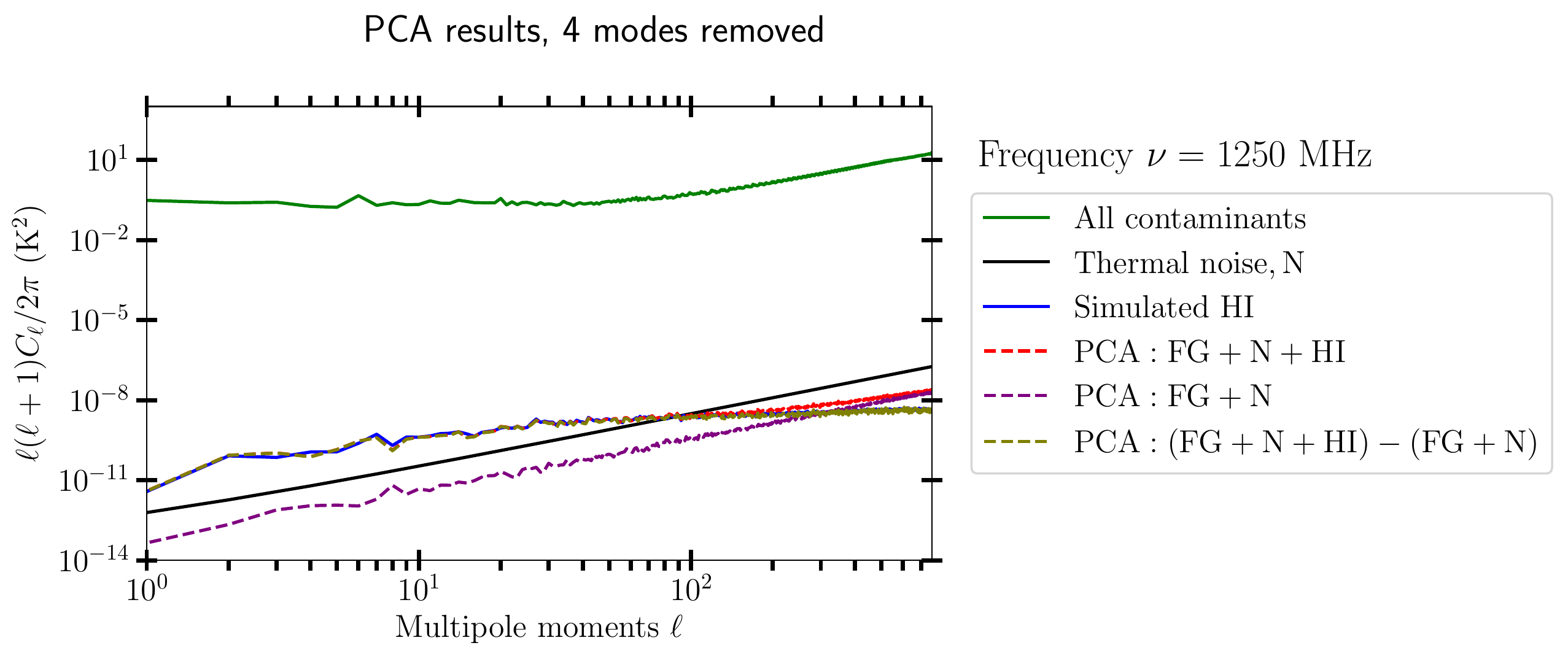}}
\caption{\hi power spectra recovery when removing different numbers of PCA modes. The map is at the frequency $1.25 \ {\rm GHz}$ and for the maximum multipole range $\ell = 768$. {\it Upper Left}--The simulations without thermal noise; 
{\it Upper Right}--The simulations with thermal noise; {\it Lower}--the simulation with instrumental noise and the de-bias subtraction.}
\label{fig:PCA_recovery_spectra}
\end{figure*}


This section is dedicated to fairly treat the foreground challenge, and provide an overview of this mammoth task which we need to undertake if we want enormous \hi IM experiments that are being put in place to succeed.

To help to maximize the scientific impact of the future 21-cm experiments, we apply PCA to model simulations as described in the previous section. We carry out PCA tests by
considering the input sky strip map (between latitudes [$25^{\circ}, 104^{\circ}$]), which is the whole region expected to be surveyed by the FAST telescope.

To realize the effect of foreground contamination, we generate and visualize the assumed superimposed sky maps for Galactic synchrotron, extragalactic point sources, thermal noise and 21-cm emission for the frequency range in which we are interested. The combined sky component models, indeed, show that the foreground contaminants whose brightness temperature are very high compared to that of \hi signal,
will overshadow the \hi signal to about $4-5$ orders of magnitude~\citep{Battye_2013, Bigot-Sazy_2015} (Fig.~\ref{contaminated_map}).

We considered the sky strip between declinations of
$[-14^{\circ}, 65^{\circ}]$ that the FAST telescope is expected to survey, by masking out unrequired patches of the sky from full simulated sky maps. We choose a single frequency channel centred at $1.25 \ {\rm GHz}$, which is the mid-range frequency in the FAST radio band, to demonstrate the foreground cleaning process applied to contaminated angular power spectra of the sky maps and visualize both the resulting power spectra and the \hi maps.

We first apply PCA to sky components without including the thermal noise; where in this case we consider the multipole moments up to $\ell = 768$ to cover the smallest angular scales that are within the FAST beam. After that, we include the thermal noise; the purpose here is to show separately how the noise power impacts the foreground subtraction process. This is because the FAST thermal noise is quite strong at small scales, and will progressively affect the \hi recovery towards larger multipoles. In the following, we discuss the results in maps and power spectra separately.

\subsection{Results of the map cleaning}

We report the results of the \hi map reconstruction by omitting the thermal noise in Fig.~\ref{fig:PCA_recovery_map}. The \hi cleaning procedure under this approach is equivalent to removing the first several principal eigenmodes because the high-order modes represent the smooth foreground modes (Fig.~\ref{smooth_foreground}). This smoothness is also illustrated in the right panel of Fig.~\ref{eigenvalues_evolution} because the first three principal axes vary with frequency smoothly. We, however, notice increasing non-linearity in the principal axes in the corresponding order of decreasing eigenvalues. The first few principal axes exhibit the property of the foreground which is smoother and more dominant than \hi signal. These principal axes pick out the dominant components projected onto the data matrix $X$.

It is clear from the results in Fig.~\ref{fig:PCA_recovery_map} that PCA can completely separate contaminants and recover \hi signal at relevant scales. As more eigenmodes being removed, the map progressively lowers its amplitude and approaches the underlying \hi signal. The left panel of Fig.~\ref{eigenvalues_evolution} gives us a clue as to how many eigenmodes we need to remove to recover the \hi signal. It also shows how much of the particular sky information is contained in each of the principal components, whereby in our case nearly $99\%$ of the dominant information is constrained within the first four principal components. Thus PCA is promising to accurately recover the signal just after removing $4$ eigenmodes.

Figure~\ref{Input_vs_Recovered_HI} shows the linear relation between the input \hi signal versus the recovered \hi signal, which shows the recovery procedure is unbiased (see also~\citet{Bigot-Sazy_2015}).
This figure provides more information on PCA performance as we consider the mean deviation between the input and the recovered \hi temperature maps,
\begin{eqnarray}
\label{min_devs}
\Delta T = \sqrt{\frac{1}{N} \sum_{i}\Big(T_{i}^{\rm in} - T_{i}^{\rm out} \Big)^{2}},
\end{eqnarray}
where $N$ is the number of pixels in the map.

We calculate the mean deviation (mean scatter) from the best fit at frequency $1250$ MHz and find $\Delta T = 0.034 \ {\rm mK}$. This deviation is slightly higher at lower frequencies because Galactic synchrotron more dominates at lower frequencies than higher frequencies. Such contamination makes the algorithm struggle to clean dirty \hi sky maps efficiently, and more contributions arguably come from high angular scales (low-$\ell$).

Next, we include the thermal noise and re-run our pipeline. The reconstructed maps in this case are very close to those in the noise-free case (Fig.~\ref{fig:PCA_recovery_map}) visually so we don't show them here. Instead, we plot and compare the \hi power spectra recovery for different PCA-mode removals in Fig.~\ref{fig:PCA_recovery_spectra}.

\subsection{Results of power spectra reconstruction}
\label{sec:2D_PS}

\begin{figure*}
\centerline{\includegraphics[width=0.45\textwidth]{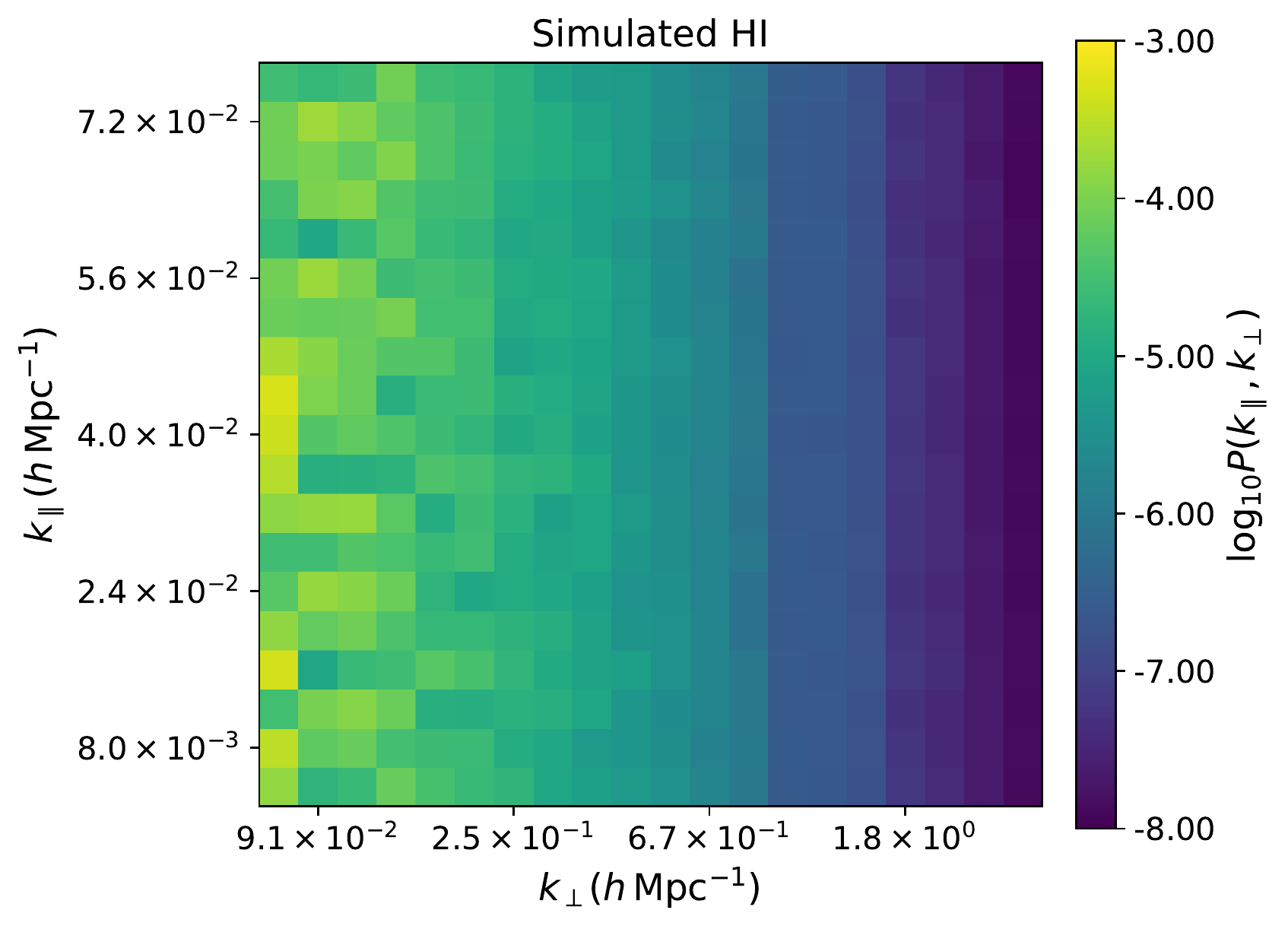}
\includegraphics[width=0.45\textwidth]{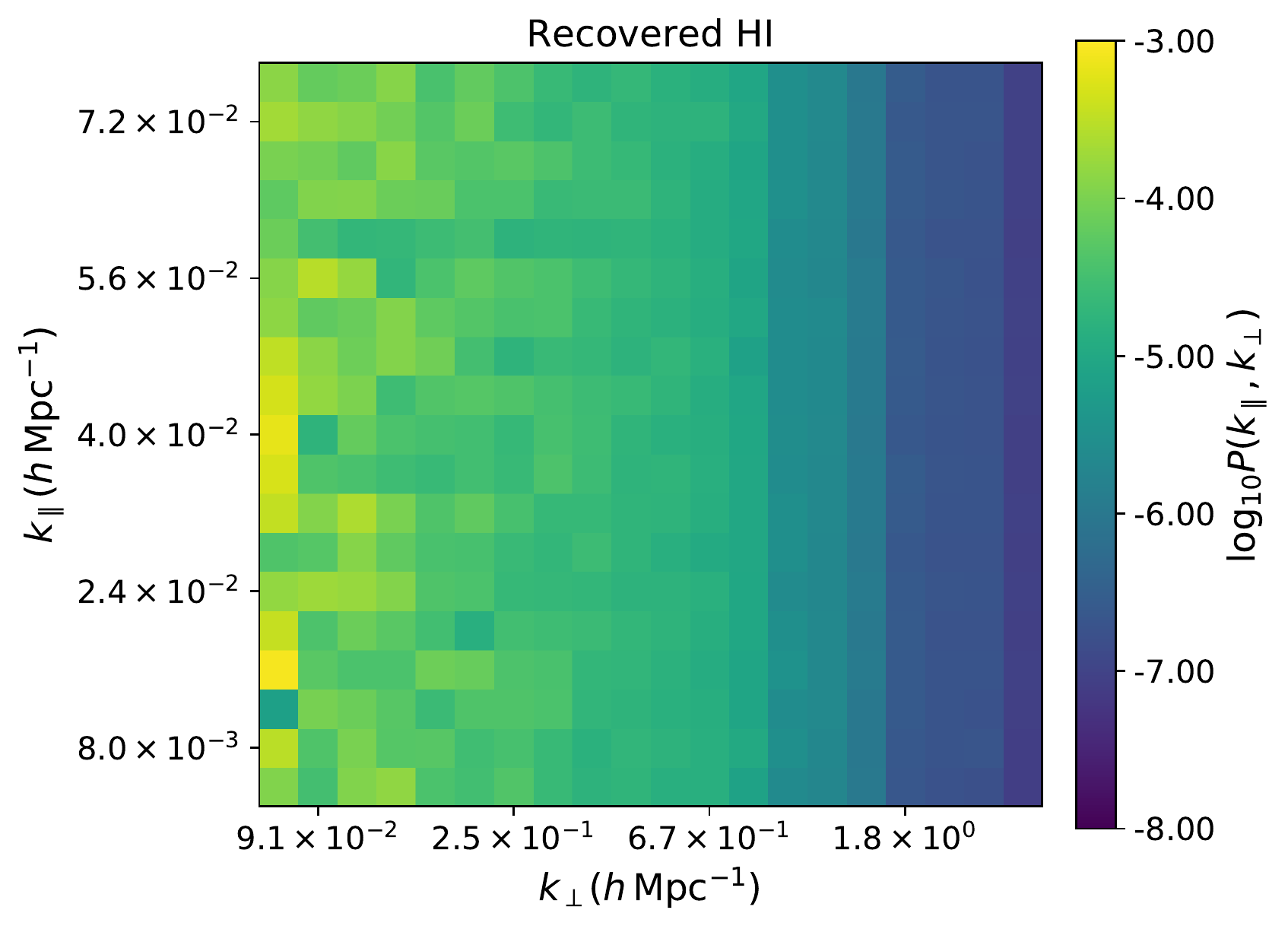}}
\centerline{\includegraphics[width=0.45\textwidth]{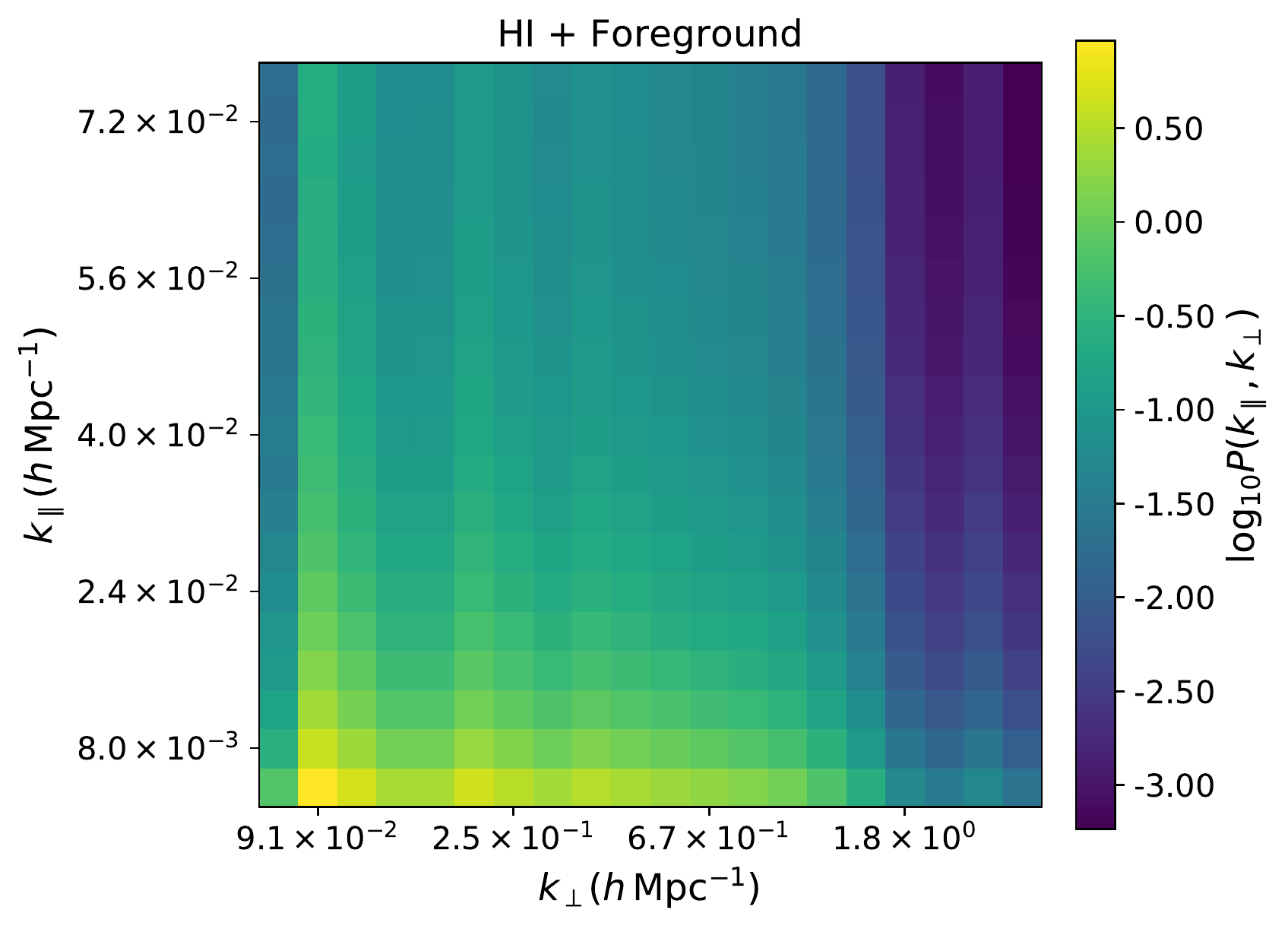}
\includegraphics[width=0.45\textwidth]{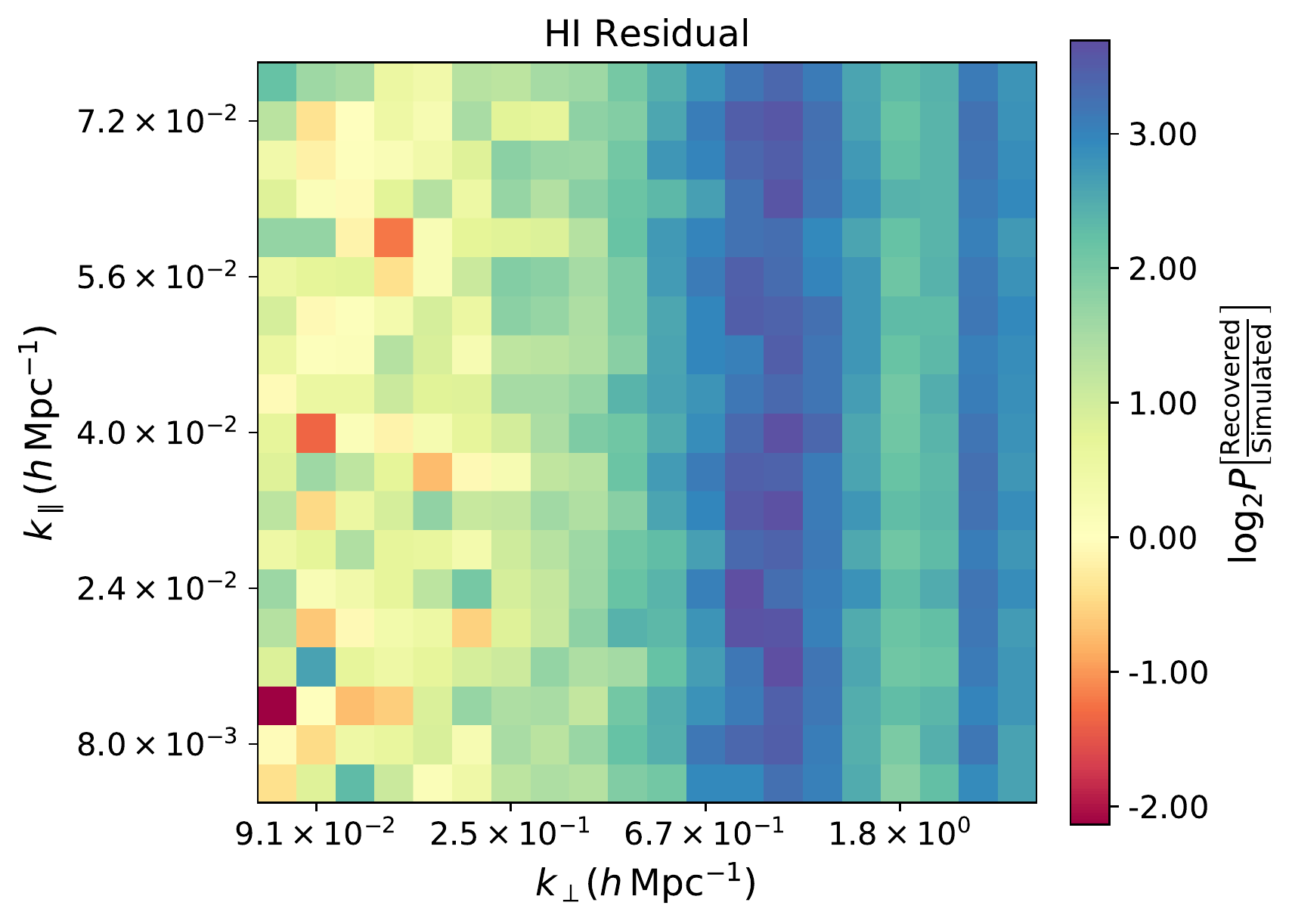}}
\caption{2D projected power spectra derived from a volume with $\nu \in[1050, 1250]$\,MHz, $\mathrm{RA}\in[-5\degree, 5\degree]$, and $\mathrm{Dec}\in [150\degree, 160\degree]$. {\it Upper left}--The simulated \hi signal; {\it Upper right}--The PCA cleaned \hi signal (4-mode removal); {\it Lower left}--The foreground contaminated signal; {\it Lower right}-- The comparison between recovered and simulated signal, defined as $\log_{2}\left(P_{\rm Recovered}/P_{\rm Simulated}\right)$.}
\label{fig:2D_PS}
\end{figure*}

\subsubsection{Power spectra in $\ell$-space}

We now calculate the power spectra recovery for different PCA-removal cases. In the left panel of Fig.~\ref{fig:PCA_recovery_spectra}, one can see that for the noise-free case, after four-modes removal, the PCA-cleaned power spectra are consistent with the underlying \hi power spectra very well and there is no bias in this reconstruction.

In the right panel of Fig.~\ref{fig:PCA_recovery_spectra}, we plot the case with the instrumental noise which includes both thermal and $1/f$ noise. We see that the progressive increase in the noise amplitude towards small angular scales causes the less accurate reconstruction of the \hi power spectrum (see also ~\citet{GNILC}). Under this consideration, beyond $\ell = 150$, the 4-mode removal power spectra have larger amplitude than the underlying \hi signal (comparing red dashed and blue curves). The reason is that beyond this point, instrumental noise increases more quickly than \hi signal. Unlike foreground emission which are coherent across frequencies, thermal noise and \hi are less coherent and have more fluctuations across-frequencies, thus the algorithm cannot easily identify the thermal noise and strip it out. 

Although various noises and systematics complicate the foreground removal across all frequencies, with known problems at high and small angular scales, the situation may become much more complicated. Therefore, in order to remove the bias of this kind, we calculate the foreground removal of the foreground plus noise only (purple dashed line in the lower panel of Fig.~\ref{fig:PCA_recovery_spectra}) and subtract it from that of foreground+noise+\hi (red dashed line in the same panel), i.e. performing the following operation
\begin{eqnarray}
\hat{C}^{\rm de-bias}_{\ell}=\hat{C}^{\rm PCA}_{\ell}\left({\rm FG+N}+\hi\right)-\hat{C}^{\rm PCA}_{\ell}\left({\rm FG+N}\right).
\end{eqnarray}
We plot the result as olive dashed line in the lower panel of Fig.~\ref{fig:PCA_recovery_spectra}. One can see that with this operation, the noise bias can be removed fairly well, and the resultant power spectrum is consistent with the input \hi power spectrum very well.

It is possible to reduce the thermal noise by altering some FAST telescope survey parameters, such as decreasing the survey area and increasing the total observational time. But we decide not to do so to account for the largest possible survey region under a reasonable optimal observational time and other
parameters. Therefore, the above study which includes the significant impact of the thermal noise is very close to reality.

\subsubsection{Power spectra in $k$-space}

We now compare our 21-cm data cube before and after foreground removal in $k$-space to see the effect of PCA. To work in $k-$space, a background model is required as we need to transform between angle and distance for which we use spatially-flat $\Lambda$CDM model. We cut a $10^{\circ}\times 10^{\circ}$ area data cube ($\mathrm{RA}\in[-5\degree, 5\degree]$, and $\mathrm{Dec}\in [150\degree, 160\degree]$) with the same frequency range $\nu \in[1050\, \mathrm{MHz}, 1250\, \mathrm{MHz}]$ from the data volume. Then we perform the Fourier transformation of this data cube to calculate the Fourier space power spectra. Therefore, we essentially neglect the evolution effect from $z = 0.35$ to $z=0.14$. Since our initial data cube is not a perfect square cube, in order to make Fourier transformation, we equalize the sizes of all slices and set the edge length at median frequency $d(\nu = 1150\, \mathrm{MHz}) = 118.8 \, h^{-1}\mathrm{Mpc}$. 

We then perform the Fourier transform based on this comoving square cubic space. The comoving distance between the emission time ($t_{\rm e}$) and observational time ($t_0$) is
\begin{eqnarray}
\chi &=& \int_{t_{\rm e}}^{t_0} c \frac{\mathrm{d} t}{a (t)} \nonumber\\
&=& \frac{c}{H_0}\int_{a_{\rm e}}^{a_0} \frac{da}{a^2 E(a)}
\end{eqnarray}
where $a(t)$ is the scale factor, $H_0$ is the Hubble constant, $E(a)=\sqrt{\Omega_{\rm m}a^{-3}+\Omega_{\Lambda}}$ is the reduced Hubble parameter. Since FAST survey occurs over a narrow radial range, the distance-frequency relation can be replaced by a linearised approximation
\begin{eqnarray}
\chi -\chi_\mathrm{ref} &=& \frac{c}{H_0}\int_{a}^{a_\mathrm{ref}} \frac{da}{a^2 E(a)} \nonumber \\
&=& \frac{c}{H_0}\int_{z_\mathrm{ref}}^{z} \frac{d z}{E(z)} \nonumber\\
&\approx& - \frac{1}{\nu_{10}}\frac{c}{H_0}\frac{(1 + z_\mathrm{ref})^2}{E(z_\mathrm{ref})} ( \nu - \nu_\mathrm{ref}) \,,
\end{eqnarray}
where $\nu$ is the frequency of a specific slice, and the subscript ``ref" denotes the reference quantity which in our case is the median frequency $1150\,{\rm MHz}$. The other assumption we adopt is the small angle approximation since we only work on the cube with $10^{\circ}$ across, so the curvature of transverse slice is ignored. The transverse distance between any two pixels in the cubic volume is $d=\chi \theta$.

We then perform the Fourier transformation on this data cube and obtain the 2D power spectra. Figure \ref{fig:2D_PS} shows the 2D projected power spectra for the simulated \hi signal (upper-left), the PCA cleaned \hi signal (upper right), the foreground contaminated signal (lower-left), and the comparison between recovered and simulated signal (lower-right), defined as $\log_{2}\left(P_{\rm recovered}/P_{\rm simulated} \right)$. If one compares the lower-left and upper-right panel, one can see that the foreground and noise modes, which correspond to the low-$k_{\parallel}$ region are systematically removed by the PCA method. As a result, the PCA removed data volume and the simulated HI signal are very close to each other in $(k_{\parallel}, k_{\perp})$ space, subject to some small mismatch at around $k_{\perp}\simeq 0.67\,h\,{\rm Mpc}^{-1}$ regime. Overall, from Fourier space analysis we confirm the effectiveness of our PCA foreground removal.

\section{Discussion and Conclusion}
\label{PCA_Anal_Conclusion}
We presented a detailed study of the Principal Component Analysis framework to subtract the foregrounds which have been
predicted to overwhelm the \hi signal. We described and analyzed some of the interconnected aspects that have been overlooked in most of the literature, and visualized both mathematical and algorithmic flow of PCA, providing a clear linkage between input and results. We show how the PCA reliably handles, processes and manipulates data, transforming it into various quantitative parametric relationships to optimize its performance and achieve the desired end. The principal component analysis can, to a large extent, redeem \hi signal from the contaminations that overshadow it. We see that, with the removal of $4$ (see Figs.~\ref{fig:PCA_recovery_map} and~\ref{fig:PCA_recovery_spectra}) principal components, we can recover the \hi signal to a very significant accuracy. Removal of fewer than this number of principal components would be possible to accurately recover the signal if the only contaminant present is Galactic synchrotron. However, because we include free-free emission, extragalactic point sources, noise components, such as thermal noise and spectrally varying $1/f$ noise, and allow synchrotron spatial/frequency variation, total foreground information becomes spread across more principal components. We find that PCA robustness increases in proportion to the strength of the frequency correlation in the frequency correlation between foreground components. The results we get are visually promising and have a virtue of unveiling the \hi cosmological information that is buried under $\sim 10^{4}$ times larger in magnitude
foreground emissions. There may however be several challenges, especially at large-angular scales (small $\ell$'s), PCA may also remove the signal itself, producing biased results of \hi. Besides, $1/f$ noise can impose challenges to PCA by complicating \hi signal-background/foreground confusion, causing the PCA algorithm to wrongly interpret $1/f$ noise which is uncorrelated in frequency as \hi signal.

It is essential to point out that the level of thermal noise will be critical for successful foreground separation and \hi recovery with PCA for FAST and other single-dish IM experiments.
We notice that, for a fixed survey area, the thermal noise level is sensitive to the observational time, and hence the integration time per pixel. Thus below some $t_{\rm pix}$ threshold, it may be difficult for PCA to recover \hi signal effectively, especially, at small angular scales, see Fig.~\ref{fig:PCA_recovery_spectra}.
%

\begin{figure}
\includegraphics[width=8.3cm]{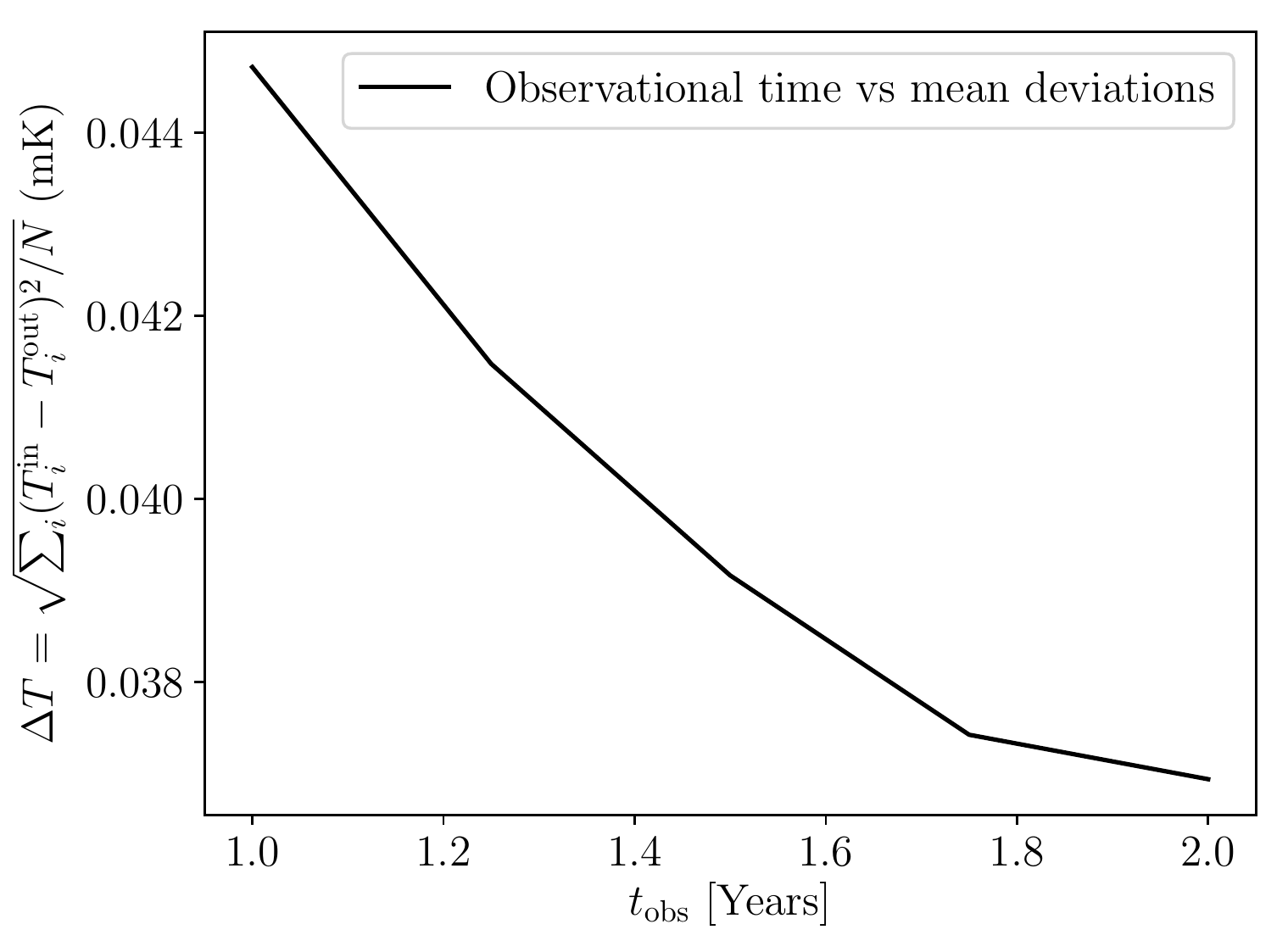}
\caption{Observational time $t_{\rm obs}$ versus the standard dispersion, $\Delta T \equiv \sqrt{\sum_{i}\left(T^{\rm in}_{i}-T^{\rm out}_{i} \right)^{2}/N}$, between input and output \hi signals.
PCA recovery precision improves with the increase in the observational time, while the rest of the experimental and survey parameters are held constant.}
\label{Mean_deviations}
\end{figure}

To illustrate this argument further, we vary the total observational time, which will automatically result in changes in the integration time per pixel and thus affect the \hi signal-to-noise ratio power.
We compute the standard deviations, $\Delta T$ for the \hi sky temperature fluctuations using Eq.~(\ref{min_devs}), where $T^{\rm in}_{i}$ and $T^{\rm out}_{i}$ are respectively, the input and output temperature fluctuations at a particular sky pixel. We plot the observational time $t_{\rm obs}$ (years) against these standard deviations as shown in Fig.~\ref{Mean_deviations}. The figure clearly illustrates how PCA \hi recovery improves (with lower variance of the reconstructed $\Delta T$) with the increase in the observational time.

Although PCA is a blind approach and assumes nothing/little about the underlying physics of the sky components, its results may unveil encoded physical information of the underlying problem.  PCA is straightforward but efficient in the sense that the data is manoeuvred and projected into axes that cleverly select an optimally parametrized frequency dependence form. The method makes essential information available in the fewest possible parameters and principal components. We feel intuitively that PCA could be more efficient and accurate if there were a mechanism in place to process data and remove the obvious artefacts. We have applied our PCA analysis to a clean simulated FAST IM data, an idealized datasets expected from the CRAFTS survey. The lessons learned in this effort can apply to SKA precursors, such as MeerKAT~\citep{MeerKLASS} and HIRAX~\citep{HIRAX},
and the final SKA datasets.

Further tests with PCA or other algorithms, will need to take into account many more artefacts, such as uncorrelated offsets, calibration errors, other systematic effects and any components which are non-signal, present in the sky maps. With consideration of such more contaminants, PCA can be hybridized with other algorithms, modified or applied in multi-stage with different methods for more reliable and robust results. Indeed, there is no hesitation for combining or testing PCA with other algorithms, since previous works from the literature have already detailed several promising approaches.

\section{Data Availability}
The pipeline that generates the simulated data is developed by our research group, and can be requested by contacting the corresponding author.

\section{Acknowledgements}

This work is supported by NSFC with grant no. 11988101, 11828301, 11633004; and NRF with grant no.105925, 109577, 120378, 120385; and the Ministry of Science and Technology Inter-government Cooperation China-South Africa Flagship program 2018YFE0120800, and Chinese Academy of Science QYZDJ-SSW-SLH017, and ``BDSS'' UKZN Research Flagship Project. We also acknowledge Drs. Yi-Chao Li and Stuart Harper for their valuable inputs and comments. E.Y. acknowledges the DAAD (German Academic Exchange Service) scholarship and the financial support from African Institute for Mathematical Sciences, University of KwaZulu-Natal, and the Dar Es Salaam University College of Education, Tanzania.

\bibliographystyle{mnras}
\bibliography{ref} 

\begin{thebibliography}{}
\makeatletter
\relax
\def\mn@urlcharsother{\let\do\@makeother \do\$\do\&\do\#\do\^\do\_\do\%\do\~}
\def\mn@doi{\begingroup\mn@urlcharsother \@ifnextchar [ {\mn@doi@}
  {\mn@doi@[]}}
\def\mn@doi@[#1]#2{\def\@tempa{#1}\ifx\@tempa\@empty \href
  {http://dx.doi.org/#2} {doi:#2}\else \href {http://dx.doi.org/#2} {#1}\fi
  \endgroup}
\def\mn@eprint#1#2{\mn@eprint@#1:#2::\@nil}
\def\mn@eprint@arXiv#1{\href {http://arxiv.org/abs/#1} {{\tt arXiv:#1}}}
\def\mn@eprint@dblp#1{\href {http://dblp.uni-trier.de/rec/bibtex/#1.xml}
  {dblp:#1}}
\def\mn@eprint@#1:#2:#3:#4\@nil{\def\@tempa {#1}\def\@tempb {#2}\def\@tempc
  {#3}\ifx \@tempc \@empty \let \@tempc \@tempb \let \@tempb \@tempa \fi \ifx
  \@tempb \@empty \def\@tempb {arXiv}\fi \@ifundefined
  {mn@eprint@\@tempb}{\@tempb:\@tempc}{\expandafter \expandafter \csname
  mn@eprint@\@tempb\endcsname \expandafter{\@tempc}}}

\bibitem[\protect\citeauthoryear{{Alonso}, D. and {Bull}, P. and {Ferreira},
  P.~G. and {Santos}, M.~G.}{{Alonso} et~al.}{2015}]{Alonso_D_2015}
{Alonso} D.,  {Bull} P.,  {Ferreira} P.~G.,   {Santos} M.~G.,  2015, \mn@doi
  [\mnras] {10.1093/mnras/stu2474}, \href
  {http://adsabs.harvard.edu/abs/2015MNRAS.447..400A} {447, 400}

\bibitem[\protect\citeauthoryear{{Bacon}, David J., et
  al.}{{Bacon}}{2020}]{SKAI_Red}
{Bacon} David~J. e.~a.,  2020, \mn@doi [\pasa] {10.1017/pasa.2019.51}, \href
  {https://ui.adsabs.harvard.edu/abs/2020PASA...37....7S} {37, e007}

\bibitem[\protect\citeauthoryear{{Banday}, A.~J. and {Wolfendale},
  A.~W.}{{Banday} \& {Wolfendale}}{1990}]{Banday_1990}
{Banday} A.~J.,  {Wolfendale} A.~W.,  1990, \mnras, \href
  {https://ui.adsabs.harvard.edu/abs/1990MNRAS.245..182B} {245, 182}

\bibitem[\protect\citeauthoryear{{Banday}, A.~J. and {Wolfendale},
  A.~W.}{{Banday} \& {Wolfendale}}{1991}]{Banday_1991}
{Banday} A.~J.,  {Wolfendale} A.~W.,  1991, \mn@doi [\mnras]
  {10.1093/mnras/248.4.705}, \href
  {https://ui.adsabs.harvard.edu/abs/1991MNRAS.248..705B} {248, 705}

\bibitem[\protect\citeauthoryear{{Battye}, Richard, et
  al.}{{Battye}}{2016}]{Battye_2016}
{Battye} Richard e.~a.,  2016, arXiv e-prints, \href
  {https://ui.adsabs.harvard.edu/abs/2016arXiv161006826B} {p. arXiv:1610.06826}

\bibitem[\protect\citeauthoryear{{Battye}, R.~A. and {Brown}, M.~L. and
  {Browne}, I.~W.~A. and {Davis}, R.~J. and {Dewdney}, P. and {Dickinson}, C.
  and {Heron}, G. and {Maffei}, B. and {Pourtsidou}, A. and {Wilkinson},
  P.~N.}{{Battye} et~al.}{2012}]{Battye_2012}
{Battye} R.~A.,  et~al., 2012, preprint, \href
  {http://adsabs.harvard.edu/abs/2012arXiv1209.1041B} {} (\mn@eprint {arXiv}
  {1209.1041})

\bibitem[\protect\citeauthoryear{{Battye}, R.~A. and {Browne}, I.~W.~A. and
  {Dickinson}, C. and {Heron}, G. and {Maffei}, B. and {Pourtsidou},
  A.}{{Battye} et~al.}{2013}]{Battye_2013}
{Battye} R.~A.,  {Browne} I.~W.~A.,  {Dickinson} C.,  {Heron} G.,  {Maffei} B.,
    {Pourtsidou} A.,  2013, \mn@doi [\mnras] {10.1093/mnras/stt1082}, \href
  {http://adsabs.harvard.edu/abs/2013MNRAS.434.1239B} {434, 1239}

\bibitem[\protect\citeauthoryear{{Bigot-Sazy}, M.-A. and {Dickinson}, C. and
  {Battye}, R.~A. and {Browne}, I.~W.~A. and {Ma}, Y.-Z. and {Maffei}, B. and
  {Noviello}, F. and {Remazeilles}, M. and {Wilkinson}, P.~N.}{{Bigot-Sazy}
  et~al.}{2015}]{Bigot-Sazy_2015}
{Bigot-Sazy} M.-A.,  et~al., 2015, \mn@doi [\mnras] {10.1093/mnras/stv2153},
  \href {http://adsabs.harvard.edu/abs/2015MNRAS.454.3240B} {454, 3240}

\bibitem[\protect\citeauthoryear{{Bigot-Sazy}, M.-A. and {Ma}, Y.-Z. and
  {Battye}, R.~A. and {Browne}, I.~W.~A. and {Chen}, T. and {Dickinson}, C. and
  {Harper}, S. and {Maffei}, B. and {Olivari}, L.~C. and {Wilkinsondagger},
  P.~N.}{{Bigot-Sazy} et~al.}{2016}]{Bigot-Sazy_2016}
{Bigot-Sazy} M.-A.,  et~al., 2016, in {Qian} L.,  {Li} D.,  eds,  Astronomical
  Society of the Pacific Conference Series Vol. 502, Frontiers in Radio
  Astronomy and FAST Early Sciences Symposium 2015. p.~41 (\mn@eprint {arXiv}
  {1511.03006})

\bibitem[\protect\citeauthoryear{Bonaldi, A. and Bedini, L. and Salerno, E. and
  Baccigalupi, C. and De Zotti, G.}{Bonaldi et~al.}{2006}]{Bonaldi_2006}
Bonaldi A.,  Bedini L.,  Salerno E.,  Baccigalupi C.,   De~Zotti G.,  2006,
  \mn@doi [Monthly Notices of the Royal Astronomical Society]
  {10.1111/j.1365-2966.2006.11025.x}, 373, 271

\bibitem[\protect\citeauthoryear{{Braun}, R. and {Bourke}, T. and {Green},
  J.~A. and {Keane}, E. and {Wagg}, J.}{{Braun} et~al.}{2015}]{SKA_science}
{Braun} R.,  {Bourke} T.,  {Green} J.~A.,  {Keane} E.,   {Wagg} J.,  2015,
  Advancing Astrophysics with the Square Kilometre Array (AASKA14), \href
  {http://adsabs.harvard.edu/abs/2015aska.confE.174B} {p.~174}

\bibitem[\protect\citeauthoryear{{Bull}, P. and {Camera}, S. and {Raccanelli},
  A. and {Blake}, C. and {Ferreira}, P. and {Santos}, M. and {Schwarz},
  D.~J.}{{Bull} et~al.}{2015a}]{SKAI}
{Bull} P.,  {Camera} S.,  {Raccanelli} A.,  {Blake} C.,  {Ferreira} P.,
  {Santos} M.,   {Schwarz} D.~J.,  2015a, Advancing Astrophysics with the
  Square Kilometre Array (AASKA14), \href
  {http://adsabs.harvard.edu/abs/2015aska.confE..24B} {p.~24}

\bibitem[\protect\citeauthoryear{{Bull}, P. and {Ferreira}, P.~G. and {Patel},
  P. and {Santos}, M.~G.}{{Bull} et~al.}{2015b}]{Bull_2015}
{Bull} P.,  {Ferreira} P.~G.,  {Patel} P.,   {Santos} M.~G.,  2015b, \mn@doi
  [\apj] {10.1088/0004-637X/803/1/21}, \href
  {http://adsabs.harvard.edu/abs/2015ApJ...803...21B} {803, 21}

\bibitem[\protect\citeauthoryear{{Chapman}, E. and {Abdalla}, F.~B. and
  {Harker}, G. and {Jeli{\'c}}, V. and {Labropoulos}, P. and {Zaroubi}, S. and
  {Brentjens}, M.~A. and {de Bruyn}, A.~G. and {Koopmans}, L.~V.~E.}{{Chapman}
  et~al.}{2012}]{blind_ICA}
{Chapman} E.,  et~al., 2012, \mn@doi [\mnras]
  {10.1111/j.1365-2966.2012.21065.x}, \href
  {http://adsabs.harvard.edu/abs/2012MNRAS.423.2518C} {423, 2518}

\bibitem[\protect\citeauthoryear{{Cunnington}, S. and {Harrison}, I. and
  {Pourtsidou}, A. and {Bacon}, D.}{{Cunnington}
  et~al.}{2019}]{Cunnington_2018}
{Cunnington} S.,  {Harrison} I.,  {Pourtsidou} A.,   {Bacon} D.,  2019, \mn@doi
  [\mnras] {10.1093/mnras/sty2928}, \href
  {https://ui.adsabs.harvard.edu/abs/2019MNRAS.482.3341C} {482, 3341}

\bibitem[\protect\citeauthoryear{{Datta}, Kanan K. and {Choudhury}, T. Roy and
  {Bharadwaj}, Somnath}{{Datta} et~al.}{2007}]{Datta_2007}
{Datta} K.~K.,  {Choudhury} T.~R.,   {Bharadwaj} S.,  2007, \mn@doi [\mnras]
  {10.1111/j.1365-2966.2007.11747.x}, \href
  {https://ui.adsabs.harvard.edu/abs/2007MNRAS.378..119D} {378, 119}

\bibitem[\protect\citeauthoryear{{Di Matteo}, Tiziana and {Perna}, Rosalba and
  {Abel}, Tom and {Rees}, Martin J.}{{Di Matteo} et~al.}{2002}]{Di_Matteo_2002}
{Di Matteo} T.,  {Perna} R.,  {Abel} T.,   {Rees} M.~J.,  2002, \mn@doi [\apj]
  {10.1086/324293}, \href
  {https://ui.adsabs.harvard.edu/abs/2002ApJ...564..576D} {564, 576}

\bibitem[\protect\citeauthoryear{{Dickinson},
  C.}{{Dickinson}}{2014}]{2014arXiv1405.7936D}
{Dickinson} C.,  2014, preprint, \href
  {http://adsabs.harvard.edu/abs/2014arXiv1405.7936D} {} (\mn@eprint {arXiv}
  {1405.7936})

\bibitem[\protect\citeauthoryear{{Dickinson}, C. and {Davies}, R.~D. and
  {Davis}, R.~J.}{{Dickinson} et~al.}{2003}]{Dickinson_2003}
{Dickinson} C.,  {Davies} R.~D.,   {Davis} R.~J.,  2003, \mn@doi [\mnras]
  {10.1046/j.1365-8711.2003.06439.x}, \href
  {https://ui.adsabs.harvard.edu/abs/2003MNRAS.341..369D} {341, 369}

\bibitem[\protect\citeauthoryear{{Gleser}, L. and {Nusser}, A. and {Benson},
  A.~J.}{{Gleser} et~al.}{2008}]{line_of_sight2}
{Gleser} L.,  {Nusser} A.,   {Benson} A.~J.,  2008, \mn@doi [\mnras]
  {10.1111/j.1365-2966.2008.13897.x}, \href
  {http://adsabs.harvard.edu/abs/2008MNRAS.391..383G} {391, 383}

\bibitem[\protect\citeauthoryear{{Hall}, A. and {Bonvin}, C. and {Challinor},
  A.}{{Hall} et~al.}{2013}]{Hall_2013}
{Hall} A.,  {Bonvin} C.,   {Challinor} A.,  2013, \mn@doi [\prd]
  {10.1103/PhysRevD.87.064026}, \href
  {http://adsabs.harvard.edu/abs/2013PhRvD..87f4026H} {87, 064026}

\bibitem[\protect\citeauthoryear{{Harper}, S.~E. and {Dickinson}, C. and
  {Battye}, R.~A. and {Roychowdhury}, S. and {Browne}, I.~W.~A. and {Ma}, Y.-Z.
  and {Olivari}, L.~C. and {Chen}, T.}{{Harper} et~al.}{2018}]{Harper_2018}
{Harper} S.~E.,  {Dickinson} C.,  {Battye} R.~A.,  {Roychowdhury} S.,  {Browne}
  I.~W.~A.,  {Ma} Y.-Z.,  {Olivari} L.~C.,   {Chen} T.,  2018, \mn@doi [\mnras]
  {10.1093/mnras/sty1238}, \href
  {http://adsabs.harvard.edu/abs/2018MNRAS.478.2416H} {478, 2416}

\bibitem[\protect\citeauthoryear{{Haslam}, C.~G.~T. and {Salter}, C.~J. and
  {Stoffel}, H. and {Wilson}, W.~E.}{{Haslam} et~al.}{1982}]{Haslam_1982}
{Haslam} C.~G.~T.,  {Salter} C.~J.,  {Stoffel} H.,   {Wilson} W.~E.,  1982,
  \aaps, \href {https://ui.adsabs.harvard.edu/abs/1982A%26AS...47....1H} {47,
  1}

\bibitem[\protect\citeauthoryear{{Hu}, Wenkai and {Wang}, Xin and {Wu},
  Fengquan and {Wang}, Yougang and {Zhang}, Pengjie and {Chen}, Xuelei}{{Hu}
  et~al.}{2020}]{Wenkai_Hu_2019}
{Hu} W.,  {Wang} X.,  {Wu} F.,  {Wang} Y.,  {Zhang} P.,   {Chen} X.,  2020,
  \mn@doi [\mnras] {10.1093/mnras/staa650}, \href
  {https://ui.adsabs.harvard.edu/abs/2020MNRAS.493.5854H} {493, 5854}

\bibitem[\protect\citeauthoryear{{Kovetz}, Ely D., et
  al.}{{Kovetz}}{2017}]{Kovetz_2017}
{Kovetz} Ely~D. e.~a.,  2017, arXiv e-prints, \href
  {https://ui.adsabs.harvard.edu/abs/2017arXiv170909066K} {p. arXiv:1709.09066}

\bibitem[\protect\citeauthoryear{{Lewis}, Antony and {Challinor},
  Anthony}{{Lewis} \& {Challinor}}{2007}]{Lewis_2007}
{Lewis} A.,  {Challinor} A.,  2007, \mn@doi [\prd]
  {10.1103/PhysRevD.76.083005}, \href
  {https://ui.adsabs.harvard.edu/abs/2007PhRvD..76h3005L} {76, 083005}

\bibitem[\protect\citeauthoryear{{Li}, Di and {Pan}, Zhichen}{{Li} \&
  {Pan}}{2016}]{Li_2016}
{Li} D.,  {Pan} Z.,  2016, \mn@doi [Radio Science] {10.1002/2015RS005877},
  \href {https://ui.adsabs.harvard.edu/abs/2016RaSc...51.1060L} {51, 1060}

\bibitem[\protect\citeauthoryear{{Li}, D. and {Wang}, P. and {Qian}, L. and
  {Krco}, M. and {Jiang}, P. and {Yue}, Y. and {Jin}, C. and {Zhu}, Y. and
  {Pan}, Z. and {Nan}, R. and {Dunning}, A.}{{Li} et~al.}{2018}]{Li_2018}
{Li} D.,  et~al., 2018, \mn@doi [IEEE Microwave Magazine]
  {10.1109/MMM.2018.2802178}, \href
  {https://ui.adsabs.harvard.edu/abs/2018IMMag..19..112L} {19, 112}

\bibitem[\protect\citeauthoryear{{Li}, Di and {Dickey}, John M. and {Liu},
  Shu}{{Li} et~al.}{2019}]{Li2019}
{Li} D.,  {Dickey} J.~M.,   {Liu} S.,  2019, \mn@doi [Research in Astronomy and
  Astrophysics] {10.1088/1674-4527/19/2/16}, \href
  {https://ui.adsabs.harvard.edu/abs/2019RAA....19...16L} {19, 016}

\bibitem[\protect\citeauthoryear{{Liu}, A. and {Tegmark}, M.}{{Liu} \&
  {Tegmark}}{2011}]{Liu_and_Tegmark_2011}
{Liu} A.,  {Tegmark} M.,  2011, \mn@doi [\prd] {10.1103/PhysRevD.83.103006},
  \href {http://adsabs.harvard.edu/abs/2011PhRvD..83j3006L} {83, 103006}

\bibitem[\protect\citeauthoryear{{Masui}, K.~W., et
  al.}{{Masui}}{2013a}]{Masui_2013}
{Masui} K.~W. e.~a.,  2013a, \mn@doi [\apjl] {10.1088/2041-8205/763/1/L20},
  \href {http://adsabs.harvard.edu/abs/2013ApJ...763L..20M} {763, L20}

\bibitem[\protect\citeauthoryear{{Masui}, K.~W., et al.}{{Masui}}{2013b}]{GBT}
{Masui} K.~W. e.~a.,  2013b, \mn@doi [\apjl] {10.1088/2041-8205/763/1/L20},
  \href {http://adsabs.harvard.edu/abs/2013ApJ...763L..20M} {763, L20}

\bibitem[\protect\citeauthoryear{{Nan}, R. and {Li}, D. and {Jin}, C. and
  {Wang}, Q. and {Zhu}, L. and {Zhu}, W. and {Zhang}, H. and {Yue}, Y. and
  {Qian}, L.}{{Nan} et~al.}{2011}]{Nan_2011}
{Nan} R.,  et~al., 2011, \mn@doi [International Journal of Modern Physics D]
  {10.1142/S0218271811019335}, \href
  {http://adsabs.harvard.edu/abs/2011IJMPD..20..989N} {20, 989}

\bibitem[\protect\citeauthoryear{{Newburgh}, L.~B., et
  al.}{{Newburgh}}{2016}]{HIRAX}
{Newburgh} L.~B. e.~a.,  2016, in Ground-based and Airborne Telescopes VI. p.
  99065X (\mn@eprint {arXiv} {1607.02059}), \mn@doi{10.1117/12.2234286}

\bibitem[\protect\citeauthoryear{{Olivari}, L.~C. and {Remazeilles}, M. and
  {Dickinson}, C.}{{Olivari} et~al.}{2016}]{GNILC}
{Olivari} L.~C.,  {Remazeilles} M.,   {Dickinson} C.,  2016, \mn@doi [\mnras]
  {10.1093/mnras/stv2884}, \href
  {http://adsabs.harvard.edu/abs/2016MNRAS.456.2749O} {456, 2749}

\bibitem[\protect\citeauthoryear{{Pacholczyk},
  A.~G.}{{Pacholczyk}}{1970}]{Pacholczyk_1970}
{Pacholczyk} A.~G.,  1970, {Radio astrophysics: Nonthermal processes in
  galactic and extragalactic sources}.
A Series of books in astronomy and astrophysics, W. H. Freeman

\bibitem[\protect\citeauthoryear{{Paciga}, G. and {Chang}, T.-C. and {Gupta},
  Y. and {Nityanada}, R. and {Odegova}, J. and {Pen}, U.-L. and {Peterson},
  J.~B. and {Roy}, J. and {Sigurdson}, K.}{{Paciga} et~al.}{2011}]{SVD1}
{Paciga} G.,  et~al., 2011, \mn@doi [\mnras]
  {10.1111/j.1365-2966.2011.18208.x}, \href
  {http://adsabs.harvard.edu/abs/2011MNRAS.413.1174P} {413, 1174}

\bibitem[\protect\citeauthoryear{Peng, B and Jin, Chengjin and Wang, Qiming and
  Zhu, Lichun and Zhu, W and Zhang, Haiyan and Nan, Rendong}{Peng
  et~al.}{2009}]{Peng_2009}
Peng B.,  Jin C.,  Wang Q.,  Zhu L.,  Zhu W.,  Zhang H.,   Nan R.,  2009,
  \mn@doi [Proceedings of the IEEE] {10.1109/JPROC.2009.2013563}, 97, 1391

\bibitem[\protect\citeauthoryear{{Planck Collaboration}}{{Planck
  Collaboration}}{2014}]{Planck-collaboration}
{Planck Collaboration} 2014, \mn@doi [\aap] {10.1051/0004-6361/201321591},
  \href {http://adsabs.harvard.edu/abs/2014A\%26A...571A..16P} {571, A16}

\bibitem[\protect\citeauthoryear{{Pourtsidou}, A. and {Bacon}, D. and
  {Crittenden}, R.}{{Pourtsidou} et~al.}{2017}]{Pourtsidou_2017}
{Pourtsidou} A.,  {Bacon} D.,   {Crittenden} R.,  2017, \mn@doi [\mnras]
  {10.1093/mnras/stx1479}, \href
  {http://adsabs.harvard.edu/abs/2017MNRAS.470.4251P} {470, 4251}

\bibitem[\protect\citeauthoryear{{Pritchard}, J.~R. and {Loeb}, A.}{{Pritchard}
  \& {Loeb}}{2012}]{21cm-review}
{Pritchard} J.~R.,  {Loeb} A.,  2012, \mn@doi [Reports on Progress in Physics]
  {10.1088/0034-4885/75/8/086901}, \href
  {http://adsabs.harvard.edu/abs/2012RPPh...75h6901P} {75, 086901}

\bibitem[\protect\citeauthoryear{{Rybicki}, George B. and {Lightman}, Alan
  P.}{{Rybicki} \& {Lightman}}{1979}]{Rybicki1979}
{Rybicki} G.~B.,  {Lightman} A.~P.,  1979, {Radiative processes in
  astrophysics}

\bibitem[\protect\citeauthoryear{{Santos}, M., et al.}{{Santos}}{2015}]{SKAI2}
{Santos} M. e.~a.,  2015, Advancing Astrophysics with the Square Kilometre
  Array (AASKA14), \href {http://adsabs.harvard.edu/abs/2015aska.confE..19S}
  {p.~19}

\bibitem[\protect\citeauthoryear{{Santos}, M.~G., et
  al.}{{Santos}}{2017}]{MeerKLASS}
{Santos} M.~G. e.~a.,  2017, preprint, \href
  {http://adsabs.harvard.edu/abs/2017arXiv170906099S} {} (\mn@eprint {arXiv}
  {1709.06099})

\bibitem[\protect\citeauthoryear{{Santos}, M.~G. and {Cooray}, A. and {Knox},
  L.}{{Santos} et~al.}{2005}]{Santos_2005}
{Santos} M.~G.,  {Cooray} A.,   {Knox} L.,  2005, \mn@doi [\apj]
  {10.1086/429857}, \href {http://adsabs.harvard.edu/abs/2005ApJ...625..575S}
  {625, 575}

\bibitem[\protect\citeauthoryear{{Shaw}, J.~R. and {Sigurdson}, K. and {Pen},
  U.-L. and {Stebbins}, A. and {Sitwell}, M.}{{Shaw} et~al.}{2014}]{Shaw_2014}
{Shaw} J.~R.,  {Sigurdson} K.,  {Pen} U.-L.,  {Stebbins} A.,   {Sitwell} M.,
  2014, \mn@doi [\apj] {10.1088/0004-637X/781/2/57}, \href
  {http://adsabs.harvard.edu/abs/2014ApJ...781...57S} {781, 57}

\bibitem[\protect\citeauthoryear{{Shaw}, J. Richard and {Sigurdson}, Kris and
  {Sitwell}, Michael and {Stebbins}, Albert and {Pen}, Ue-Li}{{Shaw}
  et~al.}{2015}]{Shaw_2015}
{Shaw} J.~R.,  {Sigurdson} K.,  {Sitwell} M.,  {Stebbins} A.,   {Pen} U.-L.,
  2015, \mn@doi [\prd] {10.1103/PhysRevD.91.083514}, \href
  {https://ui.adsabs.harvard.edu/abs/2015PhRvD..91h3514S} {91, 083514}

\bibitem[\protect\citeauthoryear{{Smoot}, George F. and {Debono}, Ivan}{{Smoot}
  \& {Debono}}{2017}]{Smoot_2017}
{Smoot} G.~F.,  {Debono} I.,  2017, \mn@doi [\aap]
  {10.1051/0004-6361/201526794}, \href
  {https://ui.adsabs.harvard.edu/abs/2017A&A...597A.136S} {597, A136}

\bibitem[\protect\citeauthoryear{{Switzer}, E.~R. and {Chang}, T.-C. and
  {Masui}, K.~W. and {Pen}, U.-L. and {Voytek}, T.~C.}{{Switzer}
  et~al.}{2015}]{Switzer_2015}
{Switzer} E.~R.,  {Chang} T.-C.,  {Masui} K.~W.,  {Pen} U.-L.,   {Voytek}
  T.~C.,  2015, \mn@doi [\apj] {10.1088/0004-637X/815/1/51}, \href
  {http://adsabs.harvard.edu/abs/2015ApJ...815...51S} {815, 51}

\bibitem[\protect\citeauthoryear{{Tegmark},
  Max}{{Tegmark}}{1998}]{Tegmark_1998}
{Tegmark} M.,  1998, \mn@doi [\apj] {10.1086/305905}, \href
  {https://ui.adsabs.harvard.edu/abs/1998ApJ...502....1T} {502, 1}

\bibitem[\protect\citeauthoryear{{Tegmark}, Max and {Eisenstein}, Daniel J. and
  {Hu}, Wayne and {de Oliveira-Costa}, Angelica}{{Tegmark}
  et~al.}{2000}]{Tegmark_2000}
{Tegmark} M.,  {Eisenstein} D.~J.,  {Hu} W.,   {de Oliveira-Costa} A.,  2000,
  \mn@doi [\apj] {10.1086/308348}, \href
  {https://ui.adsabs.harvard.edu/abs/2000ApJ...530..133T} {530, 133}

\bibitem[\protect\citeauthoryear{{Villaescusa-Navarro}, F. and {Alonso}, D. and
  {Viel}, M.}{{Villaescusa-Navarro} et~al.}{2017}]{Villaescusa-Navarro}
{Villaescusa-Navarro} F.,  {Alonso} D.,   {Viel} M.,  2017, \mn@doi [\mnras]
  {10.1093/mnras/stw3224}, \href
  {http://adsabs.harvard.edu/abs/2017MNRAS.466.2736V} {466, 2736}

\bibitem[\protect\citeauthoryear{{Wang}, X. and {Tegmark}, M. and {Santos},
  M.~G. and {Knox}, L.}{{Wang} et~al.}{2006}]{Line-of-sight}
{Wang} X.,  {Tegmark} M.,  {Santos} M.~G.,   {Knox} L.,  2006, \mn@doi [\apj]
  {10.1086/506597}, \href {http://adsabs.harvard.edu/abs/2006ApJ...650..529W}
  {650, 529}

\bibitem[\protect\citeauthoryear{{Wilson}, T.~L. and {Rohlfs}, K. and
  {H{\"u}ttemeister}, S.}{{Wilson} et~al.}{2009}]{Wilson_2009}
{Wilson} T.~L.,  {Rohlfs} K.,   {H{\"u}ttemeister} S.,  2009, {Tools of Radio
  Astronomy}.
Springer-Verlag, Berlin, Germany, \mn@doi{10.1007/978-3-540-85122-6}

\bibitem[\protect\citeauthoryear{{Wolz}, L. and {Abdalla}, F.~B. and {Blake},
  C. and {Shaw}, J.~R. and {Chapman}, E. and {Rawlings}, S.}{{Wolz}
  et~al.}{2014a}]{fast_ICA}
{Wolz} L.,  {Abdalla} F.~B.,  {Blake} C.,  {Shaw} J.~R.,  {Chapman} E.,
  {Rawlings} S.,  2014a, \mn@doi [\mnras] {10.1093/mnras/stu792}, \href
  {http://adsabs.harvard.edu/abs/2014MNRAS.441.3271W} {441, 3271}

\bibitem[\protect\citeauthoryear{{Wolz}, L. and {Abdalla}, F.~B. and {Blake},
  C. and {Shaw}, J.~R. and {Chapman}, E. and {Rawlings}, S.}{{Wolz}
  et~al.}{2014b}]{FASTICA}
{Wolz} L.,  {Abdalla} F.~B.,  {Blake} C.,  {Shaw} J.~R.,  {Chapman} E.,
  {Rawlings} S.,  2014b, \mn@doi [\mnras] {10.1093/mnras/stu792}, \href
  {http://adsabs.harvard.edu/abs/2014MNRAS.441.3271W} {441, 3271}

\bibitem[\protect\citeauthoryear{{Wolz}, L. and {Blake}, C. and {Abdalla},
  F.~B. and {Anderson}, C.~M. and {Chang}, T.-C. and {Li}, Y.-C. and {Masui},
  K.~W. and {Switzer}, E. and {Pen}, U.-L. and {Voytek}, T.~C. and {Yadav},
  J.}{{Wolz} et~al.}{2015}]{FASTICA2}
{Wolz} L.,  et~al., 2015, preprint, \href
  {http://adsabs.harvard.edu/abs/2015arXiv151005453W} {} (\mn@eprint {arXiv}
  {1510.05453})

\bibitem[\protect\citeauthoryear{{Yohana}, Elimboto and {Li}, Yi-Chao and {Ma},
  Yin-Zhe}{{Yohana} et~al.}{2019}]{Yohana_2019}
{Yohana} E.,  {Li} Y.-C.,   {Ma} Y.-Z.,  2019, \mn@doi [Research in Astronomy
  and Astrophysics] {10.1088/1674-4527/19/12/186}, \href
  {https://ui.adsabs.harvard.edu/abs/2019RAA....19..186Y} {19, 186}

\bibitem[\protect\citeauthoryear{{Zhang}, L. and {Bunn}, E.~F. and {Karakci},
  A. and {Korotkov}, A. and {Sutter}, P.~M. and {Timbie}, P.~T. and {Tucker},
  G.~S. and {Wandelt}, B.~D.}{{Zhang} et~al.}{2016}]{2016ApJS..222....3Z}
{Zhang} L.,  {Bunn} E.~F.,  {Karakci} A.,  {Korotkov} A.,  {Sutter} P.~M.,
  {Timbie} P.~T.,  {Tucker} G.~S.,   {Wandelt} B.~D.,  2016, \mn@doi [\apjs]
  {10.3847/0067-0049/222/1/3}, \href
  {http://adsabs.harvard.edu/abs/2016ApJS..222....3Z} {222, 3}

\bibitem[\protect\citeauthoryear{{Zuo}, S. and {Chen}, X. and {Ansari}, R. and
  {Lu}, Y.}{{Zuo} et~al.}{2019}]{RPCA}
{Zuo} S.,  {Chen} X.,  {Ansari} R.,   {Lu} Y.,  2019, \mn@doi [\aj]
  {10.3847/1538-3881/aaef3b}, \href
  {https://ui.adsabs.harvard.edu/abs/2019AJ....157....4Z} {157, 4}

\bibitem[\protect\citeauthoryear{{de Oliveira-Costa}, A. and {Tegmark}, M. and
  {Gaensler}, B.~M. and {Jonas}, J. and {Landecker}, T.~L. and {Reich}, P.}{{de
  Oliveira-Costa} et~al.}{2008}]{de_Oliveira-Costa_2008}
{de Oliveira-Costa} A.,  {Tegmark} M.,  {Gaensler} B.~M.,  {Jonas} J.,
  {Landecker} T.~L.,   {Reich} P.,  2008, \mn@doi [\mnras]
  {10.1111/j.1365-2966.2008.13376.x}, \href
  {https://ui.adsabs.harvard.edu/abs/2008MNRAS.388..247D} {388, 247}

\makeatother
\end{thebibliography}

\bsp	
\label{lastpage}
\end{document}